\documentclass[fleqn,usenatbib]{mnras}

\usepackage{newtxtext,newtxmath}

\usepackage[T1]{fontenc}

\DeclareRobustCommand{\VAN}[3]{#2}
\let\VANthebibliography\thebibliography
\def\thebibliography{\DeclareRobustCommand{\VAN}[3]{##3}\VANthebibliography}

\usepackage{graphicx}	
\usepackage{amsmath}	

\usepackage{longtable}

\newcommand{\teff}{$T_{\rm eff}$} 
\newcommand{\logg}{$\log g$} 
\newcommand{\kms}{km s$^{-1}$}

\title[High precision abundance measurements in M 22]{The complex stellar system M 22: confirming abundance variations with high precision differential measurements}

\author[M. McKenzie et al.]{
M. McKenzie,$^{1,2}$\thanks{E-mail: madeleine.mckenzie@anu.edu.au}
D. Yong,$^{1,2}$
A. F. Marino,$^{3,4}$
S. Monty,$^{1,2}$
E. Wang,$^{1,2}$
A. I. Karakas,$^{5,2}$
\newauthor
A. P. Milone,$^{6,3}$
M. V. Legnardi,$^{6}$
I. U.\ Roederer,$^{7,8}$
S. Martell,$^{9,2}$
D. Horta,$^{10}$\\
$^{1}$Research School of Astronomy \& Astrophysics, Australian National University, Canberra, ACT 2611, Australia, \\
$^{2}$ARC Centre of Excellence for Astrophysics in Three Dimensions (ASTRO-3D), Canberra 2611, Australia'\\
$^{3}$Istituto Nazionale di Astrofisica — Osservatorio Astronomico di Padova, Vicolo dell’Osservatorio 5, I-35122 Padua, Italy, \\
$^{4}$Istituto Nazionale di Astrofisica — Osservatorio Astrofisico di Arcetri, Largo Enrico Fermi, 5, I-50125 Firenze, Italy,\\
$^{5}$School of Physics \& Astronomy, Monash University, Clayton VIC 3800, Australia, \\
$^{6}$Dipartimento di Fisica e Astronomia “Galileo Galilei,” Università di Padova, Vicolo dell’Osservatorio 3, I-35122 Padua, Italy,\\ 
$^{7}$Department of Astronomy, University of Michigan,
1085 S.\ University Ave., Ann Arbor, MI 48109, USA \\
$^{8}$Joint Institute for Nuclear Astrophysics -- Center for the
Evolution of the Elements (JINA-CEE), USA\\
$^{9}$School of Physics, University of New South Wales, Sydney, NSW 2052, Australia\\
$^{10}$Astrophysics Research Institute, Liverpool John Moores University, 146 Brownlow Hill, Liverpool L3 5RF, UK
}

\date{Accepted 2022 July 27. Received 2022 July 17; in original form 2022 May 31}
\pubyear{2022}

\begin{document}
\label{firstpage}
\pagerange{\pageref{firstpage}--\pageref{lastpage}}
\maketitle

\begin{abstract}
M 22 (NGC 6656) is a chemically complex globular cluster-like system reported to harbour heavy element abundance variations. However, the extent of these variations and the origin of this cluster is still debated. In this work, we investigate the chemical inhomogeneity of M 22 using differential line-by-line analysis of high-quality (R = 110,000, S/N = 300 per pixel at 514 nm) VLT/UVES spectra of six carefully chosen red giant branch stars. By achieving abundance uncertainties as low as $\sim$0.01 dex ($\sim$2 per cent), this high-precision data validates the results of previous studies and reveals variations in Fe, Na, Si, Ca, Sc, Ti, Cr, Mn, Co, Ni, Zn, Y, Zr, La, Ce, Nd, Sm and Eu. Additionally, we can conﬁrm that the cluster hosts two stellar populations with a spread of at least 0.24 dex in [Fe/H] and an average s-process abundance spread of 0.65 dex. In addition to global variations across the cluster, we also ﬁnd non-negligible variations within each of the two populations, with the more metal-poor population hosting larger spreads in elements heavier than Fe than the metal-rich. We address previous works which do not identify anomalous abundances and relate our ﬁndings to our current dynamical understanding of the cluster. Given our results, we suggest that M 22 is either a nuclear star cluster, the product of two merged clusters, or an original building block of the Milky Way.
\end{abstract}




\section{Introduction}

    Deciphering the formation mechanisms of globular clusters (GCs) is a major challenge for both theorists and observers. Many early efforts to characterise these stellar systems assumed that they consisted of simple stellar populations with a single age, helium abundance and overall metallicity. However, there are always exceptions to every rule. One such exception is the enigmatic cluster M 22, which happens to be one of the first clusters to be studied in detail (\citealt{Shapley_1930_M22}; \citealt{Sawyer_1944}; \citealt{Arp_Melbourne1959}). Historically, M 22 has often been compared to $\omega$ Centauri ($\omega$ Cen; \citealt{Shapley_1930}); a probable dwarf galaxy nucleus candidate (\citealt{Bekki_Freeman2003}). However M 22 was considered to be a less extreme version of the cluster (\citealt{Hesser+1976}; \citealt{Hesser+1977}). \citealt{Hesser_Harris_1979} builds upon \citealt{LloydEvans1978} and describes four different factors which contributed to this hypothesis.

    1. For stars lying on or near the red giant branch, David Dunlap Observatory (DDO) photometry showed a wide range of both metal and CN indices (for example, for M 22 see \cite{Norris_Freeman1983}, and for $\omega$ Cen see \citealt{Freeman_Rodgers1975}, \citealt{Norris_Bessell1975} or \citealt{LloydEvans1977}). This is a manifestation of what we now understand to be the multiple stellar population phenomenon (for reviews on this topic, see \citealt{Kraft_1994}, \citealt{Gratton+2012}, \citealt{Bastian_Lardo2018} and \citealt{Gratton+2019}). 
    
    2. Both clusters were found to contain Barium stars (\citealt{Mallia_1976_omegaCen}; \citealt{Mallia_1976_M22}) and probable CH stars (\citealt{Hesser+1977}; \citealt{Harding_1962}; \citealt{Bond_1975}). At the time, no other clusters were known to contain these anomalous stars. Today, these represent \textit{s}-process enhanced populations which have since emerged in several clusters (\citealt{Norris_DaCosta1995}; \citealt{Smith+2000}; \citealt{Yong_Grundahl2008}; \citealt{Carretta+2011}; \citealt{Carretta+2013}; \citealt{Yong+2014}; \citealt{Marino+2015}). \textit{s}-process elements refer to elements heavier than iron synthesised via the slow neutron capture process, which occurs in the He-shells of low-mass asymptotic giant branch (AGB) stars. (e.g. \citealt{Clayton+1961}; \citealt{Busso+1999}; \citealt{Karakas_Lattanzio_2014}; \citealt{Kobayashi+2020}).
    
    3. The colour magnitude diagram (CMD) of M 22 appeared to show similar characteristics to the ``wide giant-branch" phenomenon identified in $\omega$ Cen (\citealt{Woolley+1966}; \citealt{Cannon_Stobie_1973}; \citealt{Bessell_Norris1976}). Later, \cite{Marino+2009} and \cite{Piotto+2012} also found a broadened sub-giant branch, indicative of stellar populations with different C+N+O abundances or different ages (\citealt{Cassisi+2008}). While significant reddening was excluded as a possible cause in $\omega$ Centauri \citep{Cannon_1980}, M 22 is situated at the edge of the dense star clouds of Sagittarius near the Galactic plane (\citealt{Shapley_Duncan1922}) and thus it is unsurprising that \cite{Richter+1999} found differential reddening in the direction of M 22. However despite this, \cite{Monaco+2004} could not rule out the presence of a small metallicity spread.

    4. There appeared to be flattening of both clusters which was correctly interpreted as rotation of the cluster (early observations include \cite{Lindsay_1956} with supporting theoretical models from \cite{King_1961}). Internal rotation within GCs has since been confirmed by several studies such as \cite{Bianchini+2013}, \cite{Lardo+2015}; \cite{Kamann+2018}, \cite{Bianchini+2018} and \cite{Cordoni+2020}. Recently, \cite{Cordoni+2020_M22_omegaCen} combined multiband photometry from the Hubble Space Telescope (HST) and ground-based facilities with Gaia Data Release 2 and HST proper motions to analyse the kinematics of both $\omega$ Cen and M 22. Both clusters share many kinematic qualities; stellar populations with different metallicities share similar motions, ellipticity and have rotation patterns with similar phases and amplitudes.
    
    This naturally leads to the suggestion that, like $\omega$ Cen, M 22 is the nucleus of a dwarf galaxy (e.g. \citealt{Bekki_Freeman2003}). This small, but growing, class of clusters, known as "Type II" clusters \citep{Milone+2017MSP}, exhibits dispersions in metallicity and/or \textit{s}-process element abundances and includes but is not limited to $\omega$ Cen, Terzan 5, M54 and M2 (\citealt{Norris_DaCosta1995}; \citealt{McKenzie_Bekki2018}; \citealt{Carretta+2010_m54}; \citealt{Yong+2014_m2}). This is juxtaposed by "Type I" GCs which exhibit homogeneous heavy element abundances \citep{Carretta+2009}. However, emerging research demonstrates that even some Type I GCs exhibit metallicity variation amongst their primordial population as inferred from photometry and spectroscopy (\citealt{Yong+2013}; \citealt{Marino+2019_3201}; \citealt{Legnardi+2022}).

    \begin{table*}
    \centering
    \renewcommand{\arraystretch}{0.8}
        \caption{Publications on M 22 abundances and whether they report an iron or other light element abundance spread within the cluster. We do not guarantee that this is an exhaustive list, and we do not include works which approach this problem from a theoretical perspective. Any references to tables relate to the publication listed in that row.}
        \label{tab:publications}
        \begin{tabular}{l l l l}
            \hline
            Publication & Fe spread? & Other element spread? & Comments\\
            \hline \hline
            \cite{Manduca_Bell_1978} & No & No & Ca abundance of RR Lyrae variables.\\
            & & & Data from \cite{Butler+1973}.\\ \hline
            \cite{Peterson_1980} & Yes, 0.56 dex difference & Not discussed & Four stars in the sample. \\ \hline
            \cite{Cohen_1981} & No & 0.8 dex difference in Na & Three stars in the sample. \\ \hline
            \cite{Gratton_1982} & No & No & Three stars in the sample. \\ \hline
            \cite{Pilachowski+1982} & Yes, 0.5 dex difference & Yes & Six stars in the sample. \\ \hline
            \cite{Norris_Freeman1983} & Not discussed & Ca, CH and CN variations & Using photometry. \\ \hline
            \cite{Frogel+1983} & Not discussed & “Tremendous range” in CN & Using infrared photometry. \\ \hline
            \cite{Wallerstein+1987} & Sample size too small & Al and Ti & Two stars. \\
            & (0.2 dex between stars) & & \\ \hline
            \cite{Gratton_Ortolani1989} & Yes, See Table 5 & Yes, see abundances in Table 5 & Three stars. \\ \hline
            \cite{Brown+1990} & Yes, 0.3 dex with non uniform & Notable N spread & Six stars in the sample. Stellar parameters\\ 
            & reddening, 0.4 dex with & & originating from \cite{Frogel+1983},\\ 
            & uniform reddening & & \cite{Cudworth_1986} \& \cite{Norris_Freeman1983}\\ \hline
            \cite{Lehnert+1991} & Yes, 0.31 dex, see Table 5 & Yes, [Ca/H] correlated with [Na/H] & 10 stars, using Ca triplet. \\ \hline
            \cite{Laird+1991} & No & Not discussed & 26 giant stars. \\ \hline
            \cite{Brown_Wallerstein1992}& Yes, 0.19 dex & Yes, see Table 5 & Spectra used is a subset of \cite{Brown+1990}.\\
            & & & Values from Table 5 CN strong and weak stars. \\ \hline
            \cite{Anthony-Twarog+1995} & No (“a spread of less than & Confirm correlation between Ca & Using uvbyCa photometry. \\ 
            & 0.2 dex cannot be excluded”)& and CN/CH as in& \\ 
            & & \cite{Norris_Freeman1983} & \\ \hline
            \cite{Richter+1999} & No & large dispersion of CN strength & Using Stromgren photometry. \\ \hline
            \cite{Monaco+2004} & No, (maximum allowed spread & Not discussed & Wide field photometry. \\
            & of $\Delta$[Fe/H] $\simeq$ 0.1-0.2 dex) & &Emphasises reddening in the cluster. \\ \hline
            \cite{Marino+2009} & Yes, average difference of 0.15 dex & Yes, see Table 6 & 17 stars in the sample. \\ \hline
            \cite{DaCosta+2009} & Yes, IQR of metal rich& Yes, Ca & 55 candidate red giants in the field. \\ 
            & and poor populations is 0.24 dex & & Based on Ca triplet. \\ \hline
            \cite{Lee+2009}& Yes & Yes, Ca & Ca hk index of the Ca-by photometry. \\ \hline
            \cite{Marino+2011} & Yes, “substantial star-to-star & Yes, & 35 red giant stars. \\
            &metallicity scatter & $\Delta$[C+N+O/Fe]$\approx$0.13 dex& \\
            &  (-2.0 $\lesssim$ [Fe/H] $\lesssim$-1.6)” &  & \\ \hline
            \cite{Roederer+2011}& Yes & Yes, \textit{s} and \textit{r} process materials & Six RGB stars. \\ \hline
            \cite{Alves-Brito+2012} & Yes, $\Delta$[Fe/H] = 0.43 dex & Yes, large C and N abundance & 11 stars. Based on IR. \\ 
            & & spreads, 0.6 dex variation in F & \\ \hline
            \cite{Joo_Lee2013}& Yes, $\Delta$[Fe/H] = 0.25 dex & Yes, $\Delta$Y = 0.09 $\pm$ 0.04 & Population models. Best fitting model \\
            & & & age difference of 0.3 $\pm$ 0.4 Gyr.\\ \hline
            \cite{Marino_2013} & Yes, see Table 2 & Yes, see Table 2 & Seven stars on the HB, main purpose to \\
            & & & measure Ba and Na abundances.\\ \hline
            \cite{DOrazi+2013} & Not discussed & Yes, F variations & Near-infrared CRIRES spectroscopic \\
            & & & observations of six cool giant stars. \\ \hline
            \cite{Gratton+2014} & Yes, see Table 5 & Yes, see Table 5 & 94 candidate stars belonging to the HB. \\ \hline
            \cite{Lim+2015}& Yes, $\Delta$[Fe/H] = 0.18 dex & Yes, Ca. Also a CN bimodality & New narrow-band Ca photometry. \\
            & & & As in \cite{Lee+2009} but without\\
            & & & contamination from CN bands.\\ \hline
            \cite{Mucciarelli+2015} & No & Yes, \textit{s}-process variations & UVES and UVES-FLAMES of 17 giants. \\ \hline
            \cite{Lee_2016} & Yes & Yes & Testing different stellar parameters. \\ \hline
            \cite{Lee_2020} & Not explicitly discussed & Yes, double CN-CH & Found five stellar populations. \\
            & & anti-correlations on the RGB& \\ \hline
            \cite{Meszaros+2020} & No & Sample too small to confirm  & 20 stars with a S/N>70 \\
            & & Ca spread. No stars to confirm & from the APOGEE survey.\\
            & &  Ce or Nd spreads & \\ \hline
            This work & Yes, >0.24 dex & Yes & 6 stars, strictly differential analysis.\\
            \hline \hline
        \end{tabular}
    \end{table*}

    Many studies suggest that M 22 is indeed a Type II cluster. However, the possibility that M 22 is a nucleated dwarf remains an unresolved issue (\citealt{Mucciarelli+2015}; \citealt{Pfeffer+2021}). 
    An extensive compilation of literature on the topic of whether M 22 contains intrinsic metallicity variations is given in Table \ref{tab:publications}. 
    
    Higher precision abundance measurements may provide new insight into the existence of a metallicity spread in M 22. Studies which reach uncertainties as low as $\sim$0.01 dex ($\sim$2 \%; \citealt{Yong+2013}) have had great success in validating abundance variations and disentangling formation sources of GCs. Therefore, the aim of this paper is to use similar high-precision abundance measurements to confirm or disprove metallicity ([Fe/H]) and other abundance variations beyond reasonable doubt. To achieve this, we reduce the uncertainties in the element abundances by adopting a differential analysis.

    In Section \ref{sec:Obs_analysis} we discuss our target selection and stellar parameter determination, we present our results from our differential abundance measurements in Section \ref{sec:results}, our discussion in Section \ref{sec:discussion} and conclusion in Section \ref{sec:conclusion}.

\section{Observations and Analysis}
\label{sec:Obs_analysis}

\subsection{Target selection and observations}
\label{sec:targets}

\begin{figure*}
    \includegraphics[width=\textwidth]{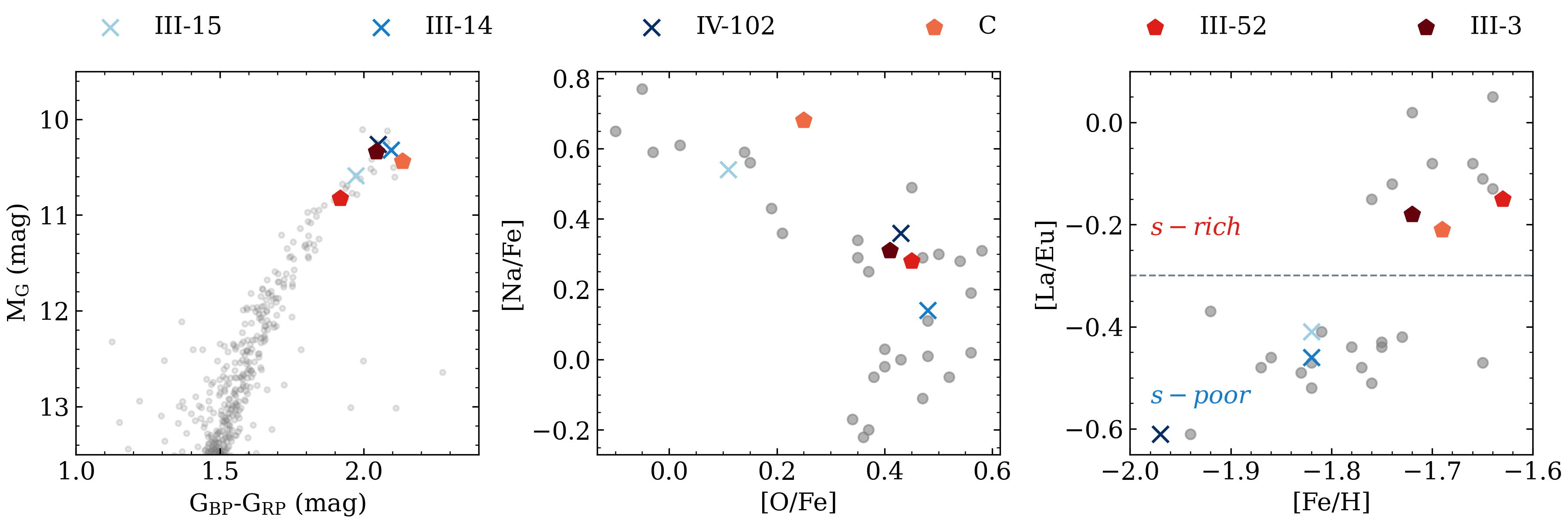}
    \caption{In each panel we show the our \textit{s}-process poor stars in blue crosses (III-15, III-14, IV-102) and \textit{s}-process rich stars in red pentagons (C, III-52, III-3). 
    \textit{Left panel:} The RGB in M 22 using Gaia filters. All targets appear at the tip of the RGB.
    \textit{Middle panel:} The Na-O anti-correlation using the sample of stars from \protect\cite{Marino+2011}. Our stars span the range of the anti-correlation. There is no distinction in light element abundances between the two \textit{s}-process groups. 
    \textit{Right panel:} The division at [La/Eu] = -0.3 between \textit{s}-process rich (top) and \textit{s}-process poor (bottom) stars as given in \protect\cite{Marino+2011}.}
    \label{fig:target_selection}
\end{figure*}

We select our program stars from \cite{Marino+2011}. The spectra analysed by \cite{Marino+2011} were visually examined to identify stars with detectable MgH molecular lines to be used for future isotopic analysis. Such analysis requires high resolution and signal-to-noise ratio (S/N) spectra, thus six stars were re-observed with UVES \citep{Dekker+2000} on the ESO VLT UT2 telescope. Three of which belong to the \textit{s}-process poor group as identified by \cite{Marino+2011}, and the remaining three are from the \textit{s}-process rich population. 

The observations were taken using image slicer \#3 and the 0\farcs3 slit. Exposure times for each star ranged from 1.5 to 2.1h. We used the 580nm setting which provided wavelength coverage from $\sim$4800\AA\ to $\sim$6800\AA\ with a small gap near 5800\AA\ due to the space between the two CCDs in the UVES camera. There were no detected neighbours within the entrance aperture (1.5×2.0 arcsec) of the image slicer which minimises contamination. The spectra were reduced using the ESO pipeline and radial velocities were estimated using IRAF. The spectra for each star has a resolution of $R$ = 110,000 and S/N $\geqslant$ 300 per pixel near 514 nm. We present measurements of the heliocentric radial velocities in Table \ref{tab:rvs} which are in agreement with literature values.

Fig. \ref{fig:target_selection} shows the red giant branch (RGB), Na-O anti-correlation and [La/Eu] as a function of [Fe/H] for the program stars. In the left panel, we use the G, BP and RP filters from Gaia EDR3 (\citealt{gaia_collab_2021}) to illustrate that our sample consists of stars exclusively located near the tip of the RGB. The brightest and faintest stars have G band magnitudes of 10.83 and 10.26, respectively. The centre and right panels both leverage the complete dataset from \cite{Marino+2011} with \textit{s}-process rich targets shown in red pentagons (C, III-52, III-3) and $s$-process poor targets in blue crosses (III-15, III-14, IV-102). The range of light element abundances for the Na-O anticorrelation is highlighted in the centre, and the distinction between the two \textit{s}-process groups is given on the left. 

\begin{figure}
    \includegraphics[width=\columnwidth]{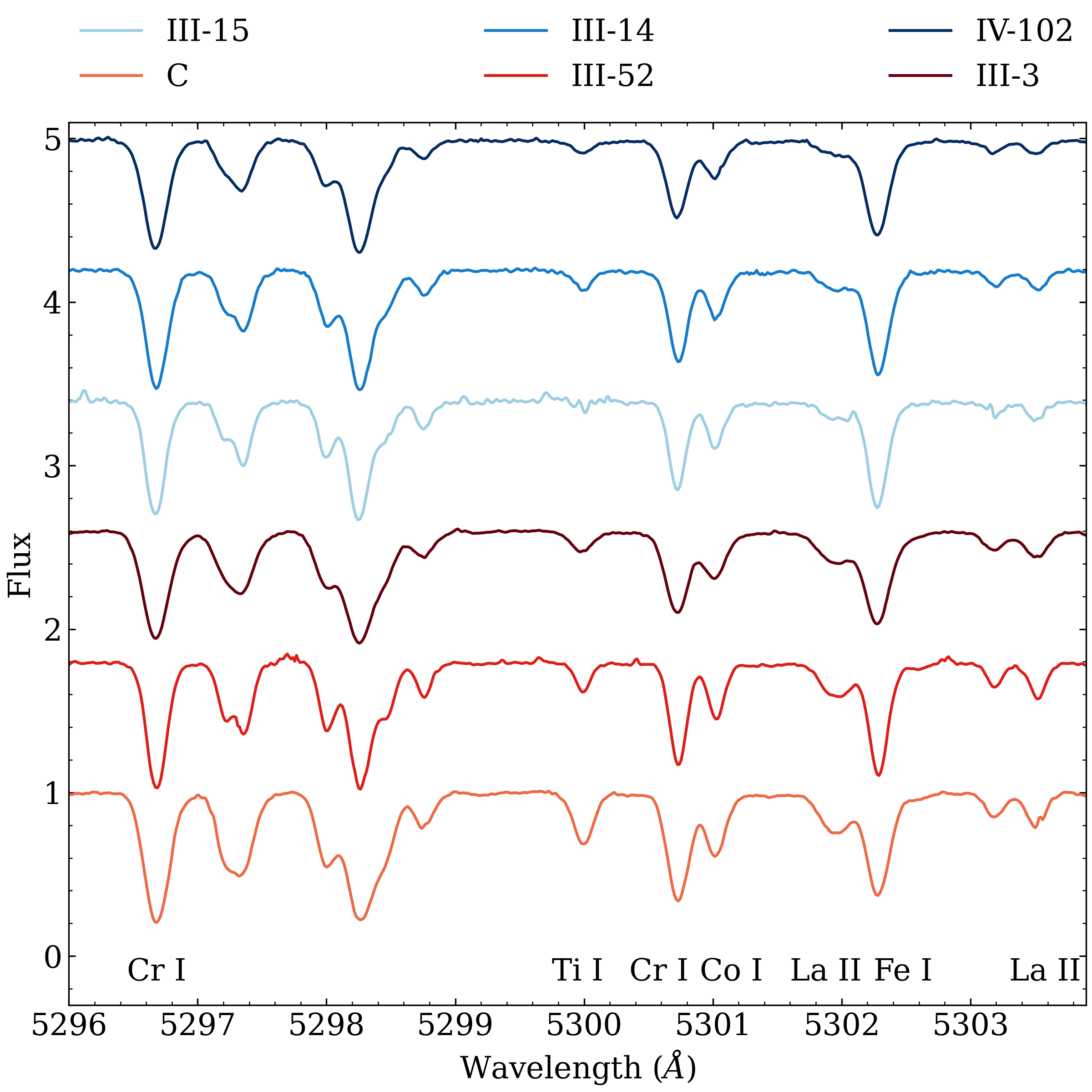}
    \caption{High signal-to-noise spectra of our target stars. Spectra increase in metallicity from the $s$-process poor stars in blue at the top to the $s$-process rich stars in red on the bottom. The blended regions from 5297 \AA{} to 5299 \AA{} are a combination of Cr I, Ti I and Fe I. The \textit{s}-process element La increases in strength from top to bottom as the metallicity increases.}
    \label{fig:spectra}
\end{figure}

We present an example of our spectra in Fig. \ref{fig:spectra} over the $s$-process line La II and iron peak elements Cr I, Ti I and Fe I. Visual inspection reveals that the $s$-process rich stars exhibit stronger
absorption lines than the $s$-process poor stars (blue). The extremely high quality of the spectra, combined with the technique of differential analysis enables us to achieve relative abundance measurements with uncertainties as low as 2\% (0.01 dex).

\subsection{Line list and equivalent width measurements}

\begin{table*}
    \centering
    \caption{Line list for our target stars and reference star (NGC 6752-mg9). The digits to the left of the decimal point in the `Species' column are the atomic number. The digit to the right of the decimal point is the ionization state (`0' = neutral, `1' = singly ionized). See the associated publications for the original references for each log \textit{gf} value. The full table is available as supplementary material. A portion is shown here for guidance regarding its form and content.}
    \label{tab:linelist}
\begin{tabular}{lllllllllll}
\hline
Wavelength & Species & L.E.P. & log \textit{gf} & C & III-3 & III-14 & III-15 & III-52 & IV-102 & NGC 6752-mg9 \\
(\AA{})      &         &(eV)    &  &(m\AA{})   &(m\AA{})&(m\AA{})&(m\AA{})&(m\AA{})&(m\AA{}) &(m\AA{})\\
\hline
5682.63 & 11.0 & 2.1 & -0.71 & 103.48 & 59.34 & 37.85 & 71.79 & 59.88 & 39.58 & 57.84 \\
5688.2 & 11.0 & 2.1 & -0.41 &  & 84.32 & 63.09 & 102.71 & 89.66 & 60.30 & 83.02 \\
6160.747 & 11.0 & 2.1 & -1.25 & 59.43 & 28.83 & 13.78 & 15.42 & 27.97 & 11.96 & 25.82 \\
5711.09 & 12.0 & 4.34 & -1.72 & 112.30 & 100.61 & 85.13 & 74.63 & 100.31 & 78.70 & 85.74 \\
6698.673 & 13.0 & 3.14 & -1.65 & 34.86 & 15.33 & 7.78 & 20.47 & 14.13 & 8.10 & 17.31 \\
\hline
\end{tabular}
\end{table*}

Our line list is an amalgamation of three lists recently used in the literature. Firstly, we use atomic data from \cite{Ji+2020streams}, which was obtained using the program \textsc{linemake}\footnote{\url{https://github.com/vmplacco/linemake}} \citep{Placco+2021}. Next, lines were also taken from \cite{Battaglia+2017} (which in turn is assembled from \citealt{Roederer+2008} and \citealt{Roederer+2010} and others) as well as the RGB tip line list used in \cite{Yong+2013}. Because of our differential approach, errors in the atomic data will largely cancel (\citealt{Melendez+2009}; \citealt{Nissen_Gustafsson2018}).

We developed and implemented a new code \textsc{review}\footnote{\url{https://github.com/madeleine-mckenzie/REvIEW}} (\textbf{R}outine for \textbf{EV}aluating and \textbf{I}nspecting \textbf{E}quivalent \textbf{W}idths) which fits individual absorption lines using the \texttt{scipy.optimize} function \texttt{curve\_fit} and a feedforward neural network (FFNN; \citealt{goodfellow2016deep}) trained on synthetic spectra. However, we do not solely rely on the FFNN as it was not trained on data that contains hyperfine and isotopic splitting. Thus we compare the output of the FFNN with a Gaussian function and select the model with the smallest residuals. We discuss this new code further in Appendix \ref{appendix:REVIEW}.

\begin{figure}
    \includegraphics[width=\columnwidth]{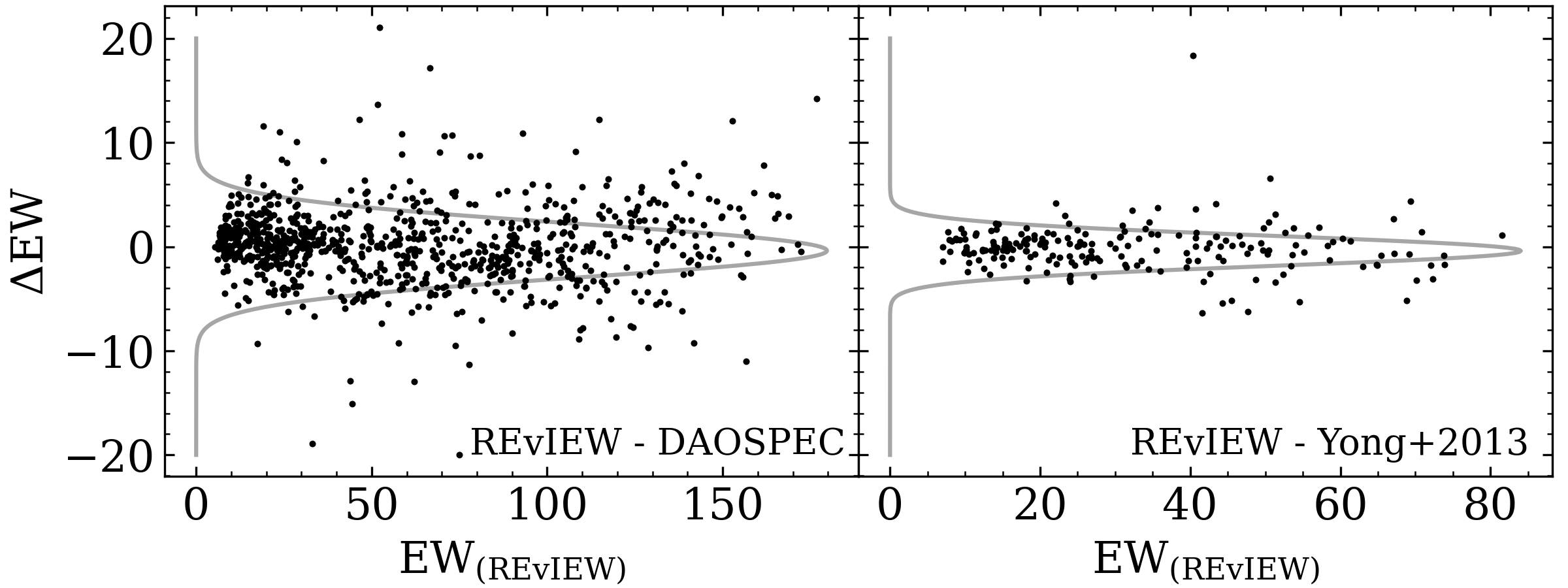}
    \caption{Comparison of EWs measured using DAOSPEC and with IRAF from \protect\cite{Yong+2013}. The left panel shows all lines measured (N = 872) using DAOSPEC for our program stars and the right panel compares EW measurements for our reference star (N = 199). Overlaid is a Gaussian fit to the distribution and we find that compared to DAOSPEC, $\mu = 0.13$ m\AA{}, $\sigma = 3.6$ m\AA{} and for \protect\cite{Yong+2013}; $\mu = -0.12$ m\AA{}, $\sigma = 2.2$ m\AA{}.}
    \label{fig:ew_error}
\end{figure}

To ensure our code is producing reliable results, we compare the equivalent widths (EWs) from \textsc{review} and DAOSpec \citep{Stetson_Pancino2008} for our program stars, and compare our measurements of our reference star with \cite{Yong+2013}. From Fig. \ref{fig:ew_error}, our method is in good agreement with both DAOSPEC and manual measurements using IRAF in \cite{Yong+2013}. As we have not manually removed poorly fit lines for this sample, EWs used to calculate abundances would be even more accurate than what each panel indicates. The advantages of our boutique method are that (i) the fitting process is much faster than IRAF, (ii) we inspect each line by eye to ensure that the continuum has been fit correctly, and (iii) we can remove blended lines or lines in low signal-to-noise regions. This is essential to remove spurious measurements in order to achieve extremely precise EW measurements. 

Strong lines, located in the flat part of the curve of growth have been excluded as their EWs are sensitive to the microturbulent velocity. 

We conservatively set this threshold to be 120 m\AA{}. Additionally, weak lines less than 5 m\AA{} have also been excluded due to larger fractional uncertainties.

\subsection{Differential stellar parameters}

\begin{table*}
\centering
    \caption{Stellar parameters and associated errors for our program stars. These stellar parameters are determined with respect to our reference star NGC 6752-mg9. For this star, we use the values from Table 1 in \protect\cite{Yong+2013}; \teff = 4288 K, \logg = 0.91 dex, [Fe/H] = -1.66 dex, $\xi$ = 1.72 \kms. We also include whether a star belongs to the \textit{s}-process rich group here for convenience. }
    \label{tab:stelar_params}
    \begin{tabular}{lllllllllc}
    \hline
Star  & \teff & $\pm$ & \logg & $\pm$  & {[}Fe/H{]} & $\pm$ & $\xi$ & $\pm$ & \textit{s}-process rich? \\ 
      & (K) & (K) &(cm s$^{-2}$) &(cm s$^{-2}$) &  &  & (km s$^{-1}$) & (km s$^{-1}$) & \\ 
\hline 
C & 3912 & 12 & 0.105 & 0.056 & -1.696 & 0.013 & 2.08 & 0.03 & \checkmark\\
III-3 & 4041 & 14 & 0.250 & 0.064 & -1.778 & 0.013 & 2.29 & 0.04 & \checkmark\\
III-14 & 4038 & 10 & 0.120 & 0.034 & -1.87 & 0.011 & 2.24 & 0.03 & $\times$\\
III-15 & 4136 & 7 & 0.450 & 0.052 & -1.825 & 0.010 & 2.03 & 0.02 & $\times$\\
III-52 & 4100 & 10 & 0.510 & 0.036 & -1.707 & 0.010 & 1.93 & 0.02 & \checkmark\\
IV-102 & 4043 & 11 & 0.100 & 0.046 & -1.973 & 0.014 & 2.43 & 0.03 & $\times$ \\   \hline          
\end{tabular}
\end{table*}

To calculate our stellar parameters, we used the program $q^2$ as detailed in \cite{Ramirez2014}. This code serves as a wrapper for the 1D LTE stellar line analysis program MOOG \citep{Sneden_1973}. $q^2$ applies a strictly differential line-by-line analysis between the program stars and a reference star by iterating over the stellar parameters in order to simultaneously minimise any correlations between (i) Fe I abundance and excitation potential $\chi$, (ii) Fe I abundance and the reduced equivalent widths ($\rm{log}(EW/\lambda)$) and (iii) removing any abundance difference between neutral (Fe I) and singly ionized iron (Fe II). Hence $q^2$ imposes excitation and ionisation balance using the abundance differences between the program stars and reference star. That is, while the absolute abundances for Fe in a given star do not necessarily achieve excitation or ionisation equilibria, the abundance differences simultaneously satisfy excitation and ionisation balance.

The model atmospheres used in the analysis were the one dimensional, plane parallel, $\alpha$ enhanced, [$\alpha$/Fe] = +0.4, NEWODF grid of ATLAS9 models by \cite{Castelli_Kurucz2003}. Other studies have shown that MARCS models yield essentially identical results to the Castelli/Kurucz models \citep{Alves_Brito+2010}. Literature values from \cite{Marino+2011} were used as our initial stellar parameters and we discuss the impact on our choice of reference star and initial parameters in Section \ref{sec:stellar_params}. For our reference star, we used NGC 6752-mg9 from \cite{Yong+2013}. This star's parameters have been estimated using photometric T$_{\rm{eff}}$ and \logg\ from isochrones which have been recommended by studies such as \cite{Mucciarelli_Bonifacio2020}. These parameters are T$_{\rm{eff}}$ = 4288K, \logg\ = 0.91 cm s$^{-2}$, [Fe/H] = -1.66 and $\xi$ = 1.72 km s$^{-1}$.

We first perform one iteration of excitation/ionization balance through $q^2$, and then remove all Fe I lines more than $1\sigma$ from the mean. We then re-run $q^2$ with this new subset of lines to obtain our final stellar parameters which are presented in Table \ref{tab:stelar_params} with their associated errors. In Table \ref{tab:lit_stelar_params}, we compare stellar parameters in the literature to our current work and find that our values are consistent with previous estimations. Although these stellar parameters are considered "spectroscopic parameters", the differential nature of the analysis means that our program stars’ iron abundances are measured with respect to our reference star on a line-by-line basis. Strict differential analysis minimises the impact of model uncertainties as well as errors in atomic data as they cancel out in each line. We further note that high-precision differential analyses deliver precise relative abundances. Such analyses have provided key breakthroughs for a range of objects including solar twins, open clusters, GCs and halo stars (e.g., \citealt{Melendez+2009}; \citealt{Bedell+2018};  \citealt{Liu+2016}; \citealt{Yong+2013}; \citealt{Reggiani+2017}). Our average errors in stellar parameters across our target stars are \teff\  $\pm$ 10.6 K, \logg\ $\pm$ 0.048 dex, [Fe/H] $\pm$ 0.012 dex and $\xi$ $\pm$ 0.03 \kms. These are significantly lower than traditional approaches (e.g. the lowest errors given in \citealt{Marino+2011} are \teff = 50 K and \logg = 0.14) and comparable to other differential studies; stars with similar stellar parameters in \cite{Yong+2013} have errors of \teff $\approx$ 20 K with \logg = 0.01 dex. In Section \ref{sec:reference_star} we discuss the outcomes of using different reference stars and guess stellar parameters.

It is known that the abundance inferred from Fe I lines suffer from non-LTE (NLTE) effects (e.g. \citealt{Thevenin_Idiart1999}; \citealt{Kraft_Ivans2003}; \citealt{Lee+2005}; \citealt{Lee_2016}) which may influence the traditional spectroscopic stellar parameter determinations. Our differential approach also minimises the impact of NLTE corrections as our reference star NGC 6752-mg9 is also an RGB tip star at a comparable temperature, gravity and metallicity to our target stars. Therefore, only differential NLTE corrections are relevant, i.e. how much does the NLTE correction change as [Fe/H] moves from -1.66 dex (reference) to -1.94 dex (most metal poor star). Similarly for T$_{\rm{eff}}$, the differences in the NLTE corrections from 4288 K (reference) to 3937 K (coolest star) are relevant. We provide examples of this differential NLTE correction in Section \ref{sec:stellar_params}.

The determination of stellar parameters for metal poor giants has been debated in the literature. \cite{Mucciarelli_Bonifacio2020} recently discussed the influence when adopting spectroscopic parameters for stars with [Fe/H] < -1.5 dex. They flag that the photometric parameters agree with best fitting stellar isochrones and thus are a more suitable choice of parameters for metal poor giants. As M 22 falls within this region, we justify our choice of stellar
parameters in Section \ref{sec:stellar_params}. While we would like to determine effective temperatures for our program stars using a photometric approach, differential reddening makes this problematic (see \citealt{Crocker_1988}, \citealt{Monaco+2004} and \citealt{Alves-Brito+2012} for reddening in the direction of M 22). Alternative photometric methods such as the infrared flux method do not achieve the required precision needed for $\sim$0.01 dex errors on our abundances (e.g. \citealt{Ramirez_Melendez_2005}). We also note that while \cite{Frebel+2013} address the issue of reconciling spectroscopic and photometric temperature scales, their analysis was applied in the regime -3.3 < [Fe/H] < -2.5 which is more metal poor than M 22.

\subsection{Chemical abundances}

Having computed our stellar parameters using a strictly differential technique, $q^2$ then calculates abundances for Fe, Na, Si, Ca, Sc, Ti, Cr, Mn, Co, Ni, Zn, Y, Zr, La, Ce, Nd, Sm and Eu. in every program star using EWs measured with \textsc{review}. We adopt the notation from \cite{Melendez+2012} where the abundance difference (program star - reference star) for a line is given by 

\[\delta A_i\ = A^{\rm{program\ star}}_{i} - A^{\rm{reference\ star}}_{i} .\] 

For a given species X, the average abundance difference is 

\[\langle \delta A^{\rm{X}}_{i} \rangle  = \frac{1}{N}\sum_{i=1}^N  \delta A^{\rm{X}}_{i} = \Delta^{\rm{X}}\ ,\]

where $N$ is the number of lines, which we write as $\Delta^{\rm{X}}$. These abundances are provided in Tables \ref{tab:chem_abundances} and \ref{tab:chem_abundances2}. We stress that these abundances are with respect to our reference star NGC 6752-mg9 from \protect\cite{Yong+2013}. We also compute these abundances on an absolute scale using the values given in Table 5 of \cite{Yong+2013}. We find that our values are in good agreement with those from \cite{Marino+2011}, with an average difference in abundance between the two studies of $\approx$0.015 dex. We find larger variations for the \textit{s}-process elements with a maximum difference of 0.14 dex between our values and abundances from \cite{Marino+2011}.

Tables \ref{tab:chem_abundances} and \ref{tab:chem_abundances2} also provide the errors on these abundance measurements. $q^2$ computes errors by adding the line-to-line scatter in quadrature with the abundance errors propagated from the stellar parameters. With an average [Fe I/H] error of 0.014, our analysis is a factor of 10 more precise when compared to stars from \cite{Marino+2011} with similar stellar parameters.

\begin{table*}
\centering
    \caption{Differential chemical abundances and their associated errors as evaluated by $q^2$ when using the reference star NGC 6752-mg9. \# represents the number of lines used to calculate the abundance. The columns $\langle \Delta^{X} \rangle$ and $\sigma \langle \Delta^{X} \rangle$ represent the average and standard deviation across all stars respectively. $\langle s\rm{-poor} \rangle$ and $\langle s\rm{-rich} \rangle$ are the average abundances divided based on their \textit{s}-process group with $\sigma_{s\rm{-poor}}$ and $\sigma_{s\rm{-rich}}$ as their accompanying standard deviations. Finally $\langle s\rm{-rich} \rangle - \langle s\rm{-rich} \rangle$ is the differences between the two \textit{s}-process groups. }
    \label{tab:chem_abundances}
\begin{tabular}{llllllllllllll}
\hline
&
  C &
  III-3 &
  III-14 &
  III-15 &
  III-52 &
  IV-102 &
  $\langle \Delta^{X} \rangle$ &
  $\sigma \langle \Delta^{X} \rangle$ &
  $\langle s\rm{-poor} \rangle$ &
  $\sigma_{s\rm{-poor}}$ &
  $\langle s\rm{-rich} \rangle$ &
  $\sigma_{s\rm{-rich}}$ &
  $\langle s\rm{-poor} \rangle - $ \\
                     &        &        &        &        &        &        &        &       &        &       &        &       & $\langle s\rm{-rich} \rangle$ \\ \hline \hline
$\Delta^{\rm{Fe I}}$  & -0.033 & -0.106 & -0.183 & -0.147 & -0.049 & -0.268 & -0.131 & 0.088 & -0.199 & 0.062 & -0.063 & 0.038 & 0.137                         \\
$\pm$                & 0.010  & 0.018  & 0.014  & 0.012  & 0.011  & 0.018  &        &       &        &       &        &       &                               \\
\#                   & 99     & 111    & 112    & 114    & 120    & 102    &        &       &        &       &        &       &                               \\\hline
$\Delta^{\rm{Fe II}}$ & -0.019 & -0.102 & -0.179 & -0.143 & -0.041 & -0.268 & -0.125 & 0.092 & -0.197 & 0.064 & -0.054 & 0.043 & 0.143                         \\
$\pm$                & 0.040  & 0.030  & 0.025  & 0.021  & 0.025  & 0.028  &        &       &        &       &        &       &                               \\
\#                   & 15     & 16     & 16     & 17     & 16     & 16     &        &       &        &       &        &       &                               \\\hline
$\Delta^{\rm{Na I}}$  & 0.315  & -0.140 & -0.417 & 0.000  & -0.077 & -0.413 & -0.122 & 0.276 & -0.277 & 0.240 & 0.033  & 0.247 & 0.309                         \\
$\pm$                & 0.059  & 0.035  & 0.017  & 0.172  & 0.017  & 0.043  &        &       &        &       &        &       &                               \\
\#                   & 2      & 3      & 3      & 3      & 3      & 3      &        &       &        &       &        &       &                               \\\hline
$\Delta^{\rm{Si I}}$  & 0.118  & 0.077  & -0.145 & -0.144 & 0.037  & -0.223 & -0.047 & 0.141 & -0.171 & 0.045 & 0.077  & 0.041 & 0.248                         \\
$\pm$                & 0.035  & 0.042  & 0.021  & 0.022  & 0.034  & 0.033  &        &       &        &       &        &       &                               \\
\#                   & 10     & 11     & 11     & 11     & 11     & 11     &        &       &        &       &        &       &                               \\\hline
$\Delta^{\rm{Ca I}}$  & 0.044  & -0.121 & -0.262 & -0.190 & 0.042  & -0.343 & -0.138 & 0.159 & -0.265 & 0.077 & -0.012 & 0.095 & 0.253                         \\
$\pm$                & 0.029  & 0.031  & 0.026  & 0.021  & 0.022  & 0.028  &        &       &        &       &        &       &                               \\
\#                   & 10     & 17     & 20     & 20     & 17     & 21     &        &       &        &       &        &       &                               \\\hline
$\Delta^{\rm{Sc II}}$ & -0.097 & -0.137 & -0.230 & -0.113 & -0.120 & -0.227 & -0.154 & 0.059 & -0.190 & 0.067 & -0.118 & 0.020 & 0.072                         \\
$\pm$                & 0.065  & 0.070  & 0.028  & 0.052  & 0.053  & 0.033  &        &       &        &       &        &       &                               \\
\#                   & 3      & 3      & 3      & 3      & 3      & 3      &        &       &        &       &        &       &                               \\\hline
$\Delta^{\rm{Ti I}}$  & 0.163  & -0.033 & -0.142 & -0.134 & 0.081  & -0.235 & -0.050 & 0.150 & -0.170 & 0.056 & 0.070  & 0.098 & 0.241                         \\
$\pm$                & 0.039  & 0.042  & 0.032  & 0.026  & 0.028  & 0.030  &        &       &        &       &        &       &                               \\
\#                   & 28     & 34     & 35     & 44     & 34     & 35     &        &       &        &       &        &       &                               \\\hline
$\Delta^{\rm{Ti II}}$ & 0.119  & 0.032  & -0.133 & -0.124 & 0.022  & -0.180 & -0.044 & 0.118 & -0.146 & 0.030 & 0.058  & 0.053 & 0.203                         \\
$\pm$                & 0.061  & 0.061  & 0.033  & 0.034  & 0.041  & 0.040  &        &       &        &       &        &       &                               \\
\#                   & 11     & 11     & 14     & 14     & 13     & 13     &        &       &        &       &        &       &                               \\\hline
$\Delta^{\rm{Cr I}}$  & 0.092  & -0.090 & -0.161 & -0.166 & 0.036  & -0.250 & -0.090 & 0.131 & -0.192 & 0.050 & 0.013  & 0.093 & 0.205                         \\
$\pm$                & 0.058  & 0.060  & 0.060  & 0.054  & 0.072  & 0.057  &        &       &        &       &        &       &                               \\
\#                   & 5      & 6      & 7      & 8      & 7      & 7      &        &       &        &       &        &       &                               \\\hline
$\Delta^{\rm{Cr II}}$ & 0.305  & 0.090  & -0.080 & -0.100 & 0.075  & -0.150 & 0.023  & 0.169 & -0.110 & 0.036 & 0.157  & 0.129 & 0.267                         \\
$\pm$                & 0.176  & 0.018  & 0.090  & 0.110  & 0.106  & 0.051  &        &       &        &       &        &       &                               \\
\#                   & 2      & 2      & 2      & 2      & 2      & 2      &        &       &        &       &        &       &                               \\\hline
$\Delta^{\rm{Mn I}}$  & 0.040  & -0.060 & -0.145 & -0.130 & 0.000  & -0.330 & -0.104 & 0.132 & -0.202 & 0.111 & -0.007 & 0.050 & 0.195                         \\
$\pm$                & 0.024  & 0.047  & 0.023  & 0.042  & 0.082  & 0.027  &        &       &        &       &        &       &                               \\
\#                   & 1      & 2      & 2      & 2      & 2      & 2      &        &       &        &       &        &       &                               \\\hline
$\Delta^{\rm{Co I}}$  & -0.020 & -0.105 & -0.225 & -0.140 & -0.030 & -0.365 & -0.148 & 0.131 & -0.243 & 0.114 & -0.052 & 0.046 & 0.192                         \\
$\pm$                & 0.000  & 0.035  & 0.045  & 0.010  & 0.030  & 0.095  &        &       &        &       &        &       &                               \\
\#                   & 1      & 2      & 2      & 2      & 2      & 2      &        &       &        &       &        &       &                               \\\hline
$\Delta^{\rm{Ni I}}$  & 0.081  & -0.033 & -0.170 & -0.130 & -0.017 & -0.249 & -0.086 & 0.119 & -0.183 & 0.061 & 0.010  & 0.062 & 0.193                         \\
$\pm$                & 0.036  & 0.023  & 0.019  & 0.016  & 0.015  & 0.023  &        &       &        &       &        &       &                               \\
\#                   & 38     & 40     & 41     & 42     & 42     & 42     &        &       &        &       &        &       &                               \\\hline
$\Delta^{\rm{Zn I}}$  & 0.110  & 0.330  & -0.070 & -0.170 & 0.040  & -0.260 & -0.003 & 0.212 & -0.167 & 0.095 & 0.160  & 0.151 & 0.327                         \\
$\pm$                & 0.023  & 0.023  & 0.015  & 0.015  & 0.014  & 0.015  &        &       &        &       &        &       &                               \\
\#                   & 1      & 1      & 1      & 1      & 1      & 1      &        &       &        &       &        &       &                               \\ \hline \hline     
\end{tabular}
\end{table*}

\begin{table*}
\centering
    \caption{An extension of Table \ref{tab:chem_abundances} but for s and \textit{r}-process elements.}
    \label{tab:chem_abundances2}
\begin{tabular}{llllllllllllll}
\hline
&
  C &
  III-3 &
  III-14 &
  III-15 &
  III-52 &
  IV-102 &
  $\langle \Delta^{X} \rangle$ &
  $\sigma \langle \Delta^{X} \rangle$ &
  $\langle s\rm{-poor} \rangle$ &
  $\sigma_{s\rm{-poor}}$ &
  $\langle s\rm{-rich} \rangle$ &
  $\sigma_{s\rm{-rich}}$ &
  $\langle s\rm{-poor} \rangle - $ \\
                     &        &        &        &        &        &        &        &       &        &       &        &       & $\langle s\rm{-rich} \rangle$ \\ \hline \hline
$\Delta^{\rm{Y II}}$  & 0.251  & 0.257  & -0.326 & -0.232 & 0.211  & -0.420 & -0.043 & 0.316 & -0.326 & 0.094 & 0.240  & 0.025 & 0.566                         \\
$\pm$                & 0.048  & 0.075  & 0.028  & 0.028  & 0.034  & 0.040  &        &       &        &       &        &       &                               \\
\#                   & 8      & 7      & 9      & 9      & 9      & 9      &        &       &        &       &        &       &                               \\\hline
$\Delta^{\rm{Zr II}}$ & 0.140  & 0.250  & -0.230 & -0.400 & 0.140  & -0.380 & -0.080 & 0.290 & -0.337 & 0.093 & 0.177  & 0.064 & 0.513                         \\
$\pm$                & 0.012  & 0.012  & 0.009  & 0.010  & 0.012  & 0.009  &        &       &        &       &        &       &                               \\
\#                   & 1      & 1      & 1      & 1      & 1      & 1      &        &       &        &       &        &       &                               \\\hline
$\Delta^{\rm{La II}}$ & 0.092  & 0.060  & -0.347 & -0.255 & 0.150  & -0.505 & -0.134 & 0.271 & -0.369 & 0.126 & 0.101  & 0.046 & 0.470                         \\
$\pm$                & 0.043  & 0.043  & 0.025  & 0.022  & 0.028  & 0.037  &        &       &        &       &        &       &                               \\
\#                   & 6      & 6      & 6      & 6      & 6      & 6      &        &       &        &       &        &       &                               \\\hline
$\Delta^{\rm{Ce II}}$ & 0.290  & 0.117  & -0.307 & -0.193 & 0.217  & -0.460 & -0.056 & 0.306 & -0.320 & 0.134 & 0.208  & 0.087 & 0.528                         \\
$\pm$                & 0.130  & 0.017  & 0.013  & 0.143  & 0.012  & 0.027  &        &       &        &       &        &       &                               \\
\#                   & 3      & 3      & 3      & 3      & 3      & 3      &        &       &        &       &        &       &                               \\\hline
$\Delta^{\rm{Nd II}}$ & 0.294  & 0.137  & -0.131 & -0.116 & 0.195  & -0.308 & 0.012  & 0.231 & -0.185 & 0.107 & 0.209  & 0.079 & 0.394                         \\
$\pm$                & 0.050  & 0.040  & 0.035  & 0.031  & 0.025  & 0.032  &        &       &        &       &        &       &                               \\
\#                   & 12     & 13     & 14     & 14     & 13     & 13     &        &       &        &       &        &       &                               \\\hline
$\Delta^{\rm{Sm II}}$ & 0.180  & 0.310  & -0.050 & 0.030  & 0.100  & -0.270 & 0.050  & 0.200 & -0.097 & 0.155 & 0.197  & 0.106 & 0.293                         \\
$\pm$                & 0.011  & 0.015  & 0.007  & 0.010  & 0.011  & 0.009  &        &       &        &       &        &       &                               \\
\#                   & 1      & 1      & 1      & 1      & 1      & 1      &        &       &        &       &        &       &                               \\\hline
$\Delta^{\rm{Eu II}}$ & -0.100 & -0.080 & -0.150 & -0.070 & -0.080 & -0.300 & -0.130 & 0.088 & -0.173 & 0.117 & -0.087 & 0.012 & 0.087                         \\
$\pm$                & 0.021  & 0.020  & 0.010  & 0.016  & 0.015  & 0.020  &        &       &        &       &        &       &                               \\
\#                   & 1      & 1      & 1      & 1      & 1      & 1      &        &       &        &       &        &       &                               \\ \hline \hline     
\end{tabular}
\end{table*}

\section{Results}
\label{sec:results}
\subsection{Differences between \textit{s}-process groups}

\begin{figure}
    \includegraphics[width=\columnwidth]{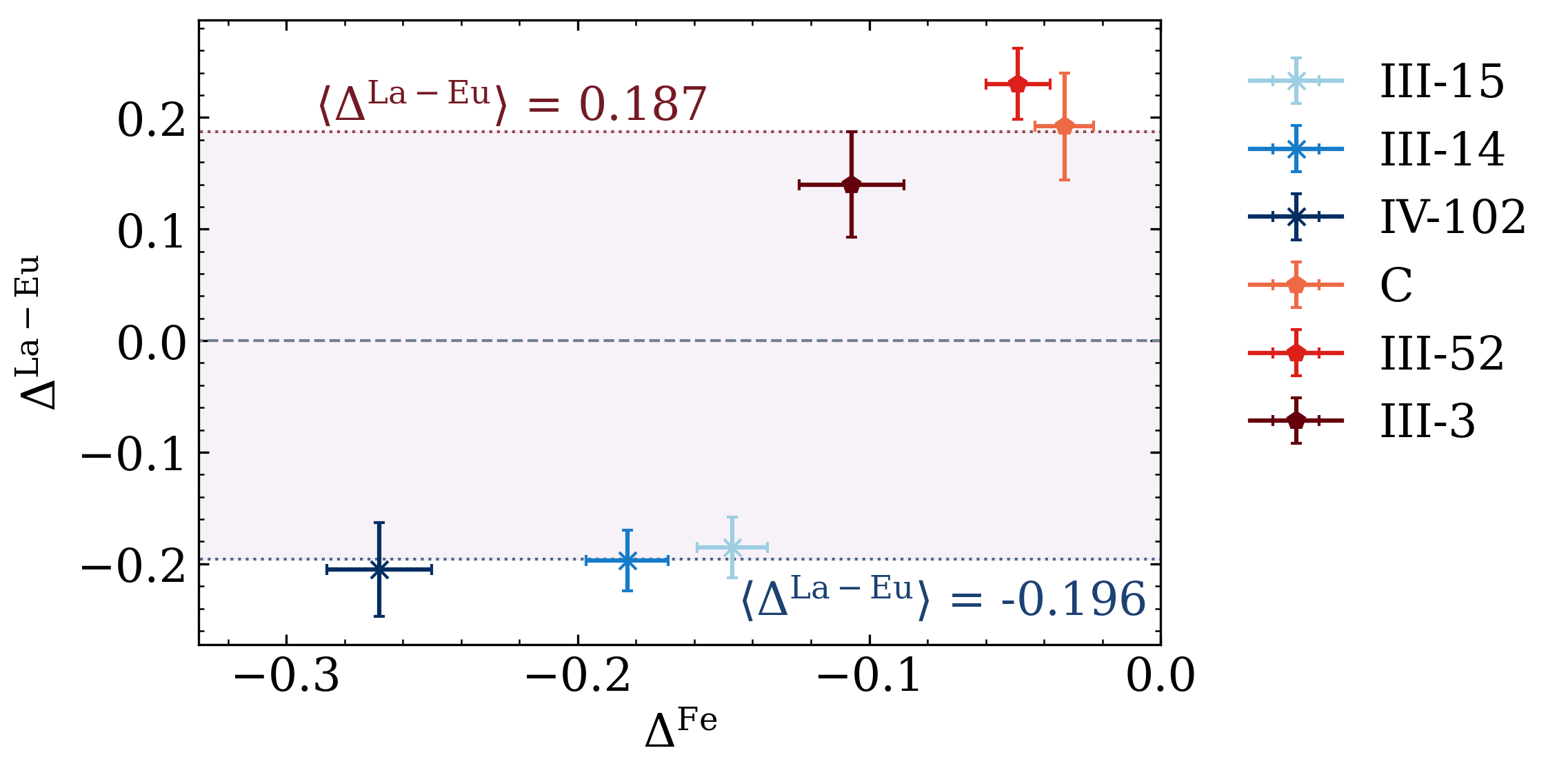}
    \caption{The $\Delta^{\rm{La-Eu}}$ abundances as a function of $\Delta^{\rm{Fe}}$ (analogous to [La/Eu] and [Fe/H] in square bracket notation respectively). Using high-precision differential abundance measurements we still recover the split between the two \textit{s}-process populations of stars. The horizontal dashed line in grey represents the $\Delta^{\rm{La-Eu}}$ of the reference star and the horizontal blue and red lines are the average values of $\Delta^{\rm{La-Eu}}$ for the \textit{s}-process poor and rich groups respectively. These results are in very good agreement with \protect\cite{Marino+2011}.}
    \label{fig:LaEu_Fe_relation}
\end{figure}

We reproduce the divide between the two \textit{s}-process groups using $\Delta^{\rm{La - Eu}}$ as a function of $\Delta^{\rm{Fe}}$ in Fig. \ref{fig:LaEu_Fe_relation} (where these two values are analogous to [La/Eu] and [Fe/H] in square bracket notation). As in \cite{Marino+2011}, we see a clear separation in our program stars and find that there must be at least a 0.24 dex iron abundance spread in the cluster as well as a spread of at least 0.38 dex for $\Delta^{\rm{La - Eu}}$. The mean iron abundance difference between the two populations is 0.14 dex. This is in very good agreement with \cite{Marino+2011} who found a difference of 0.39 dex in [La/Eu], and a mean abundance difference of 0.15 dex in Fe (see the right panel of Fig. \ref{fig:target_selection}). Therefore, we confirm the results from \cite{Lee_2016}, \cite{Marino+2011}, \cite{Roederer+2011}, \cite{DaCosta+2009} and several others, that this is an anomalous stellar system with heavy element abundance variations.

\begin{figure}
    \includegraphics[width=\columnwidth]{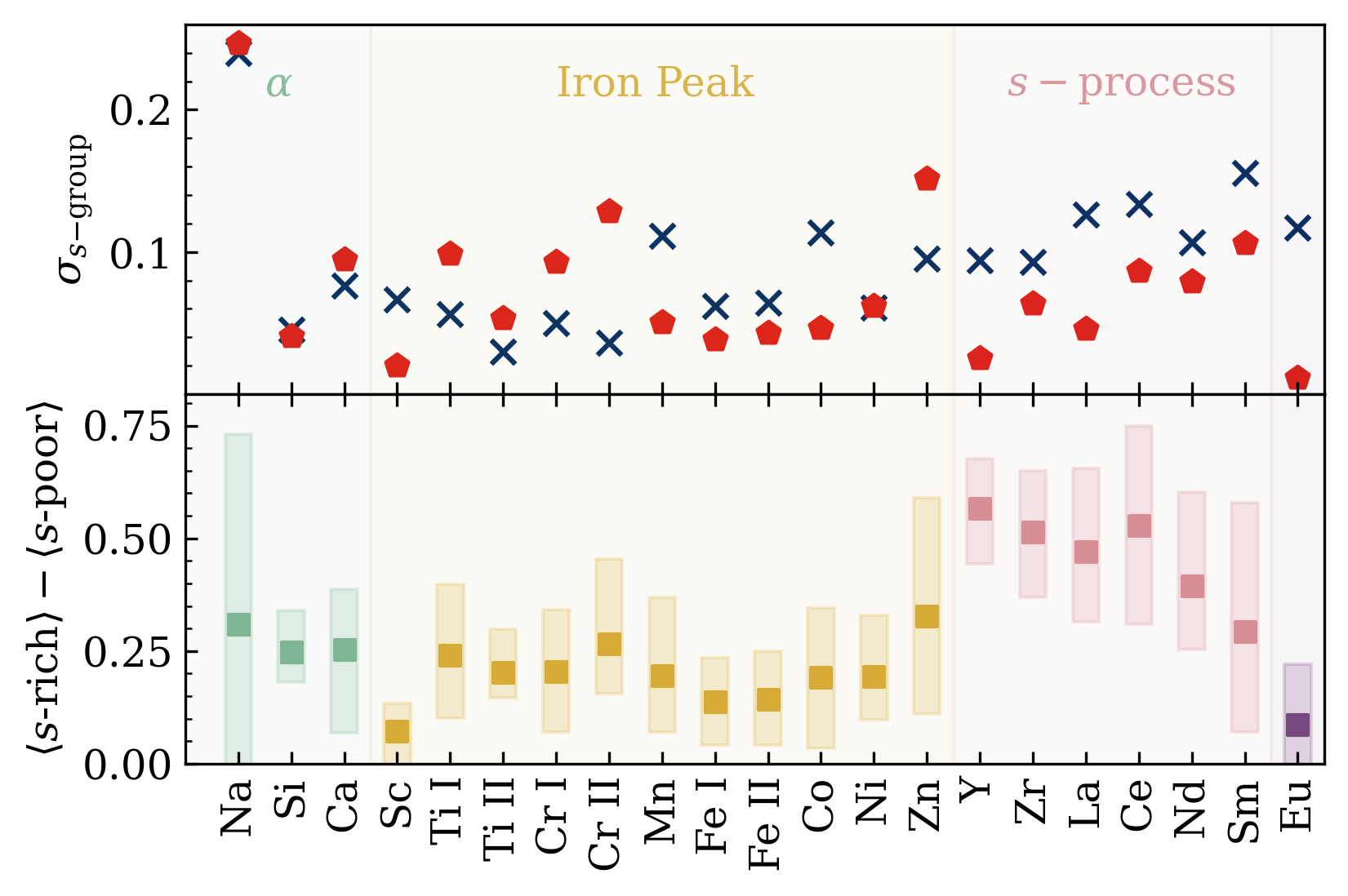}
    \caption{Comparing the different abundance variations within each \textit{s}-process group. $\alpha$-elements are shown in green, iron peak elements in yellow, \textit{s}-process elements in pink and our \textit{r}-process element, Eu, in purple. \textit{Top panel:} the $1\sigma$ dispersion for the \textit{s}-process poor group in blue crosses, and the \textit{s}-process rich group in red pentagons. The \textit{s}-process poor population has a higher dispersion across all measured \textit{s} and \textit{r}-process elements. \textit{Bottom panel:} the difference in abundances between the average \textit{s}-process poor ($\langle s-\rm{poor}\rangle$) and \textit{s}-process rich ($\langle s-\rm{rich}\rangle$) populations for each measured element. The shaded rectangle for each element represents the minimum and maximum abundance differences between the two populations and the square denotes the average difference between the two stellar groups ($|\langle s-\rm{poor} \rangle - \langle s-\rm{rich}\rangle|$), in Tables \ref{tab:chem_abundances} and \ref{tab:chem_abundances2}. The shaded rectangles for Na, Sc and Eu have overlapping abundances for the two populations so we crop their rectangles at 0.}
    \label{fig:pop_diff}
\end{figure}

Tables \ref{tab:chem_abundances} and \ref{tab:chem_abundances2} list the mean abundance and dispersion in each \textit{s}-process group for each element, along with the average difference between each group (i.e. `$\langle s\rm{-rich} \rangle - \langle s\rm{-rich} \rangle$'). These quantities are presented in Fig. \ref{fig:pop_diff} with elements colour coded by their main nucleosynthesis sites. In the top panel, the blue crosses and red pentagons represent the dispersion of the \textit{s}-process poor and rich populations respectively. For our $\alpha$-elements, each population has roughly the same dispersion with Na showing the largest spread overall due to light-element abundance variations within the cluster (as expected given the middle panel of Fig. \ref{fig:target_selection}). Iron peak elements show a mix of different dispersions, however Fe in particular has a larger variation in its \textit{s}-process poor population. Assuming this is the primordial population in the cluster, this could be a related phenomenon to the iron abundance spread in first generation of stars observed in some Galactic GCs (e.g. \citealt{Marino+2019_3201}; \citealt{Legnardi+2022}) and predicted by numerical simulations (\citealt{McKenzie_Bekki2021b}). For M 22, Fig. \ref{fig:pop_diff} also shows that all measured \textit{s}-process elements have a larger spread in the \textit{s}-process poor populations and the magnitude of these variations increases with atomic number. 

In the bottom panel, the solid square is the mean difference between the two \textit{s}-process groups and the shaded boxes are representative of the minimum and maximum range of abundances within the cluster. For Na, Sc and Eu, abundances between the two populations overlap with one another, therefore we set the minimum variation to 0. For the $\alpha$ and iron peak elements, there is an average abundance difference of $\approx$ 0.2 dex. For the \textit{s}-process elements however, as the atomic number increases, the mean separation decreases, but the total range in abundances increases. On average, there is a $\approx$ 0.5 dex difference and Y shows the largest spread between the two populations with a separation of almost 0.6 dex. These anomalous abundances represent a complex formation history and provide additional constraints to pinpoint the neutron sources and/or combination of neutron sources that are responsible for these differences. 

We emphasise however that these results are limited by our small sample size and may not be a true reflection of the abundance differences within the cluster.

\subsection{Chemical correlations}

\begin{figure}
    \includegraphics[width=\columnwidth]{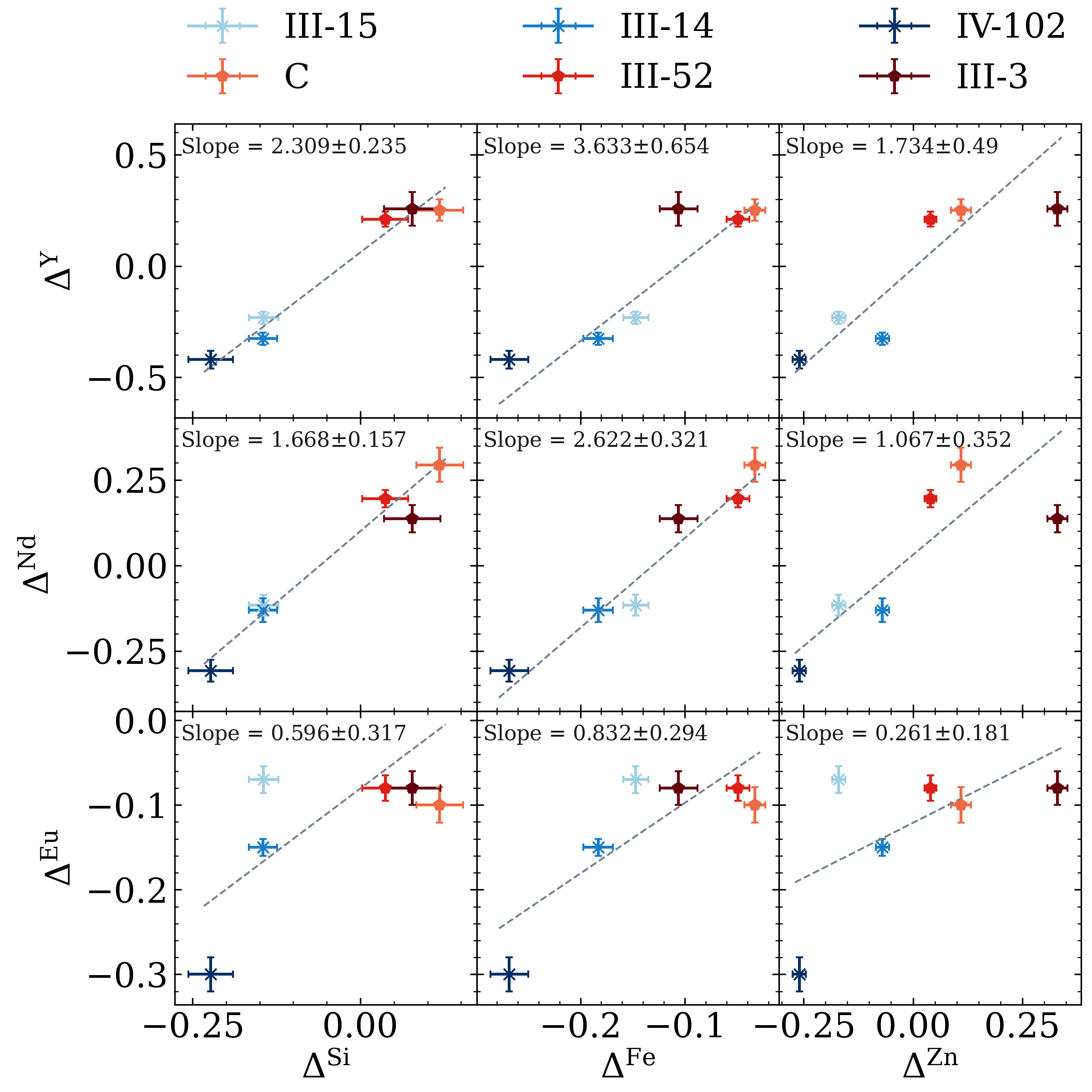}
    \caption{A grid of correlations between different elements. The $\Delta^{\rm{X}}$ notation is analogous to [X/H] in square bracket notation. On the Y-axis we include Y as a first \textit{s}-process peak element, Nd as a second \textit{s}-process peak element and Eu as an \textit{r}-process element. On the X-axis, we use Si as a proxy for alpha elements and both Fe and Zn as iron peak elements.}
    \label{fig:6_element_grid}
\end{figure}

To further examine whether or not M 22 is indeed an anomalous cluster, we take a sample of elements each representing different nucleosynthetic processes and astrophysical sources and plot them in Fig. \ref{fig:6_element_grid}. On the Y-axis, we use Y to represent a first peak \textit{s}-process element from AGB stars, Nd to represent second-peak elements and Eu to represent elements made by the \textit{r}-process. For the X-axis, Si is a proxy for alpha elements while Fe and Zn are our examples of iron-peak elements \citep{Kobayashi+2020}. Although Fe and Zn trace similar nucleosynthesis sites, Zn is produced in high energy explosions, more so than Fe. Additionally, Fe is also made in larger quantities in Type Ia supernovae compared to Zn.

For each element, we find abundance variations greater than the measurement uncertainty. The combination of Si-Y is particularly good for separating the two different \textit{s}-process populations. There are no obvious correlations for the \textit{r}-process element Eu. Additionally, it is unclear why the most metal poor star, IV-102, has an Eu abundance $\sim$0.15 dex lower than the other program stars while still having very small abundance errors. The stellar parameters for this star are in good agreement with \cite{Marino+2011} (see Table \ref{tab:lit_stelar_params}), however, this may be an artefact of our differential approach with IV-102 being 0.3 dex more metal poor than our reference star. Eu can only be measured using the 6645\AA{} line which we manually inspected to ensure accurate EW fitting, and have included in Appendix \ref{appendix:REVIEW}.

\begin{figure}
    \includegraphics[width=\columnwidth]{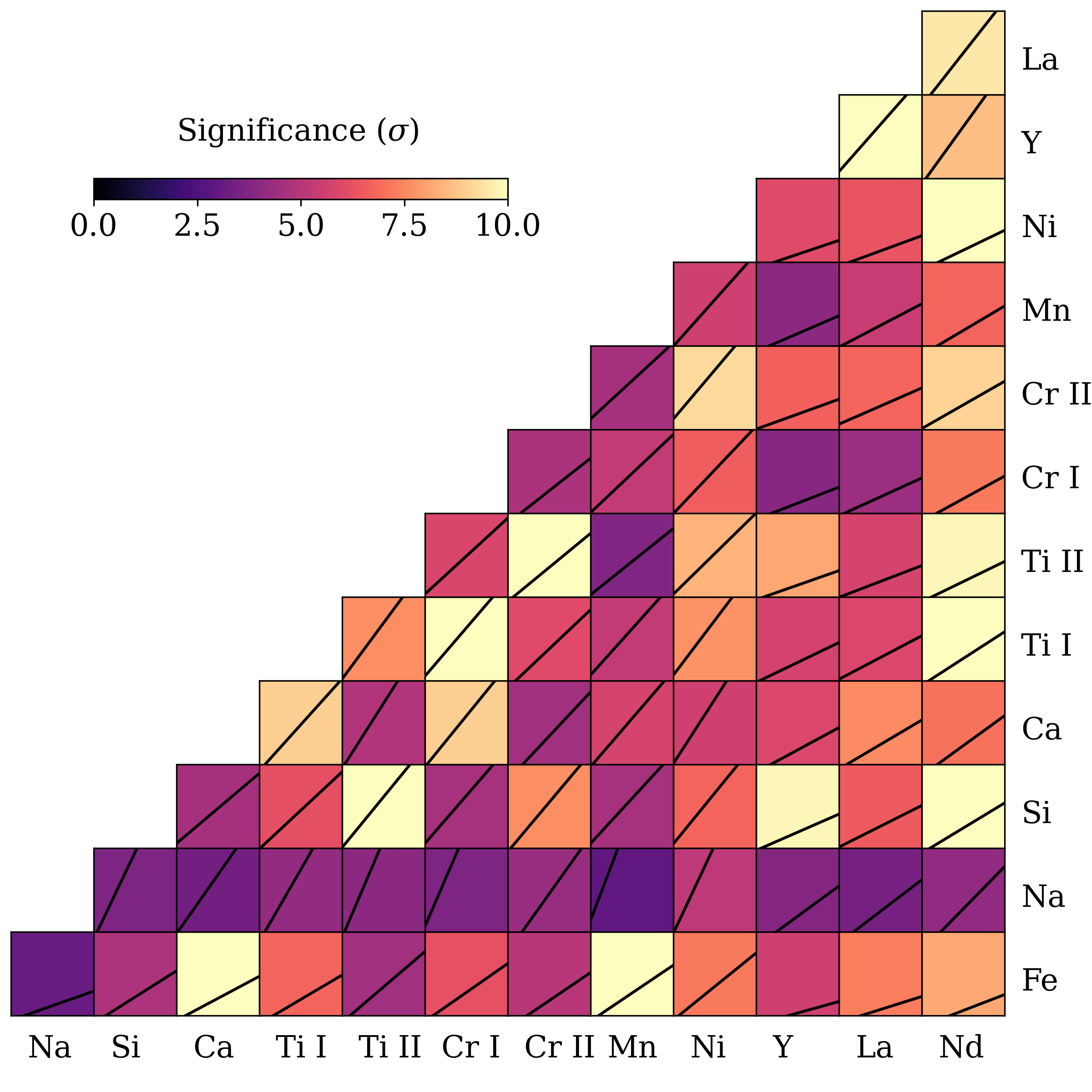}
    \caption{Linear fit to $\Delta^{\rm{X}}$ versus $\Delta^{\rm{Y}}$ (as in Fig. \ref{fig:6_element_grid}), for a subset of elements measured with at least three lines. The plots along the X and Y axis all share the same plot ranges. The significance of the gradients is indicated by the colour bar. For example, Na has a less significant correlation between other elements compared to Nd.}
    \label{fig:EVEC_plot}
\end{figure}

Taking only the linear fits from Fig. \ref{fig:6_element_grid} and applying this method to a larger sample of elements, Fig. \ref{fig:EVEC_plot} illustrates the correlations between all the elements with the significance of its correlation colouring each axis. This serves as a comparison to figures 16, 17, 19 and 20 presented in \cite{Yong+2013} and bears some similarity to the RGB bump sample used in this work. We select only elements which were measured using at least two lines. The tight correlation between Y and Si is illustrated here by $\approx$10 $\sigma$ significance. Na has correlations with lower statistical significance compared to other elements which is a result of the light element abundance variations within the cluster. As expected, we see the \textit{s}-process elements Y, La and Nd strongly correlate with each other. 

\section{Discussion}
\label{sec:discussion}
\subsection{An abundance spread in M 22?}
\label{sec:Fe_spread}

As evident from Table \ref{tab:publications}, the debate as to whether M 22 has a range in heavy element abundance has swung back and forth due to a number of divergent results. A recent work reporting no iron spread in the cluster from \cite{Mucciarelli+2015} was a prominent factor in \cite{Pfeffer+2021} rejecting M 22 as a nuclear star cluster. Our study has the highest resolution and S/N data ever taken of M 22 stars, allowing us to provide a definitive answer to the question of whether M 22 contains significant abundance variations. 

\cite{Norris+2001} introduced a `figure of merit' (F) for spectra; F = R x S/N / $\lambda$ where R = spectral resolution, S/N = signal-to-noise ratio and $\lambda$ = wavelength. The most popular spectroscopic data set of M 22 comes from \cite{Marino+2009}, which has wavelength range 4800-6800 \AA{}, a resolution R $\simeq$ 45000, and a typical S/N between 100 - 120, thus resulting in F = 1058 at $\lambda$ = 5100 \AA{}. This is a factor of six lower than our data which has R = 110,000 and has S/N=300 at 5100\AA{}, giving F = 6470. Furthermore, the high quality EW measurements and differential analysis approach employed in this work using $q^2$ enables typical relative abundance errors of 0.01 dex, and is a proven method to deliver the highest possible precision \citep{Nissen_Gustafsson2018}. \cite{Yong+2013} applied this method to NGC 6752, a cluster with previously no evidence of an \textit{s}-process or Fe abundance variations, and uncovered Fe spreads of $\sim$0.02 - 0.03 dex. This makes our study extremely well suited to discussing the intricacies of determining whether M 22 contains a genuine heavy abundance spread.

\subsubsection{Determination of stellar parameters}
\label{sec:stellar_params}

\cite{Brown+1990} stated that "Persistent but equivocal evidence of variable reddening in M 22 has confused efforts to ascertain the reality of an abundance spread in the cluster". As evident from the extinction map in \cite{Alves-Brito+2012}, reddening presents a source of uncertainty for determining stellar parameters in M 22 through various photometric methods. However \cite{Marino+2011} emphasised that spectroscopic [Fe/H] measurements do not suffer the effects of differential reddening. Conclusions from \cite{Mucciarelli+2015} act as a precursor to the discussion in \cite{Mucciarelli_Bonifacio2020} which warn against using spectroscopic parameters at metallicities lower than -1.5 dex. They find that spectroscopic parameters are inconsistent with the position of the stars in the CMD, thus resulting in an underestimate of the temperatures and gravities. Regarding the use of isochrones for this type of investigation, we note that \citet{Joyce_Chaboyer2018} provide a detailed discussion of the uncertainties in the input physics for theoretical stellar evolution models and note that careful considerations and analysis need to be taken into account. By using photometric or spectroscopic \teff\ and a photometric \logg, \cite{Mucciarelli+2015} reported no abundance spread in M 22, finding a narrow, symmetric [Fe II/H] distribution with a value of -1.75 $\pm$ 0.01 dex. They also analysed stars in NGC 6752 from \cite{Yong+2013}, including the reference star used in this work, and found a unimodal distribution for all methods of determining parameters. However, \cite{Marino_2015IAU} discusses this work in the context of C+N+O abundance differences between the two populations. When plotting the stellar parameters from \cite{Mucciarelli+2015} with isochrones which account for the CNO variations, the \logg\ values are systematically affected, but not to the same degree for the two populations; the s-poor population has systematically higher \logg.

\cite{Lee_2016} provides a comprehensive discussion of the results from \cite{Mucciarelli+2015}, echoing the conclusion from \cite{Marino_2015IAU} that incorrect surface gravities were likely used. Additionally, the metallicity of the input atmosphere models and the separation in Fe II can be amplified if different methods are used to compute these parameters. Other studies have cited that the discrepancies between different methods have been attributed to non-local thermodynamic equilibrium (NLTE) effects. \cite{Lind+2012} presents a detailed discussion of the influence of NLTE on Fe, and by extension, the impact it has on stellar parameters. They explain that the metallicity of a star is primarily based on Fe I lines which are far less sensitive to surface gravity variations compared to Fe II lines. However, these lines are subject to significant NLTE effects and the LTE ionization balance is not always realistic, resulting in an underestimation of surface gravity and metallicity. Comparing our results to previous parameters in the literature in Table \ref{tab:lit_stelar_params}, the \logg\ determined in our study are indeed lower than other works (on the order of $\sim$0.2 dex), however our metallicities agree with previous estimates.

NLTE corrections for Fe are negligible for this work. Using corrections from \cite{Lind+2012} and taking stellar parameters of program star III-3 from \cite{Mucciarelli+2015} (method 2) as an example (\teff\ = 3960k, \logg\ = 0.34, [Fe/H] = -1.8), the NLTE corrections are $\sim$0.01 dex using the closest matching parameters in their grid. Furthermore, as we are using a differential approach, it is the "differential NLTE" corrections that will impact our abundances, which by design, will also be negligible. Taking a sample of Fe lines from the INSPECT database\footnote{Data obtained from the INSPECT database, version 1.0 (www.inspect-stars.net).} (\citealt{Bergemann+2012}; \citealt{Lind+2012}), the difference in NLTE corrections between the reference and our most metal poor star, IV-102, was on the order of 0.001 dex.

\subsubsection{Choice of reference star}
\label{sec:reference_star}

\cite{Mucciarelli_Bonifacio2020} cite inadequacies of the adopted physics, in particular the assumption of 1-dimensional geometry, as the origin of the diverging spectroscopic and photometric parameters. As we have shown in previous sections, our approach of using differential measurements addresses this concern as this method aims to minimise errors arising from 1D models between the program and reference stars (for a review on differential methods, see \citealt{Nissen_Gustafsson2018}).

\begin{figure}
    \includegraphics[width=\columnwidth]{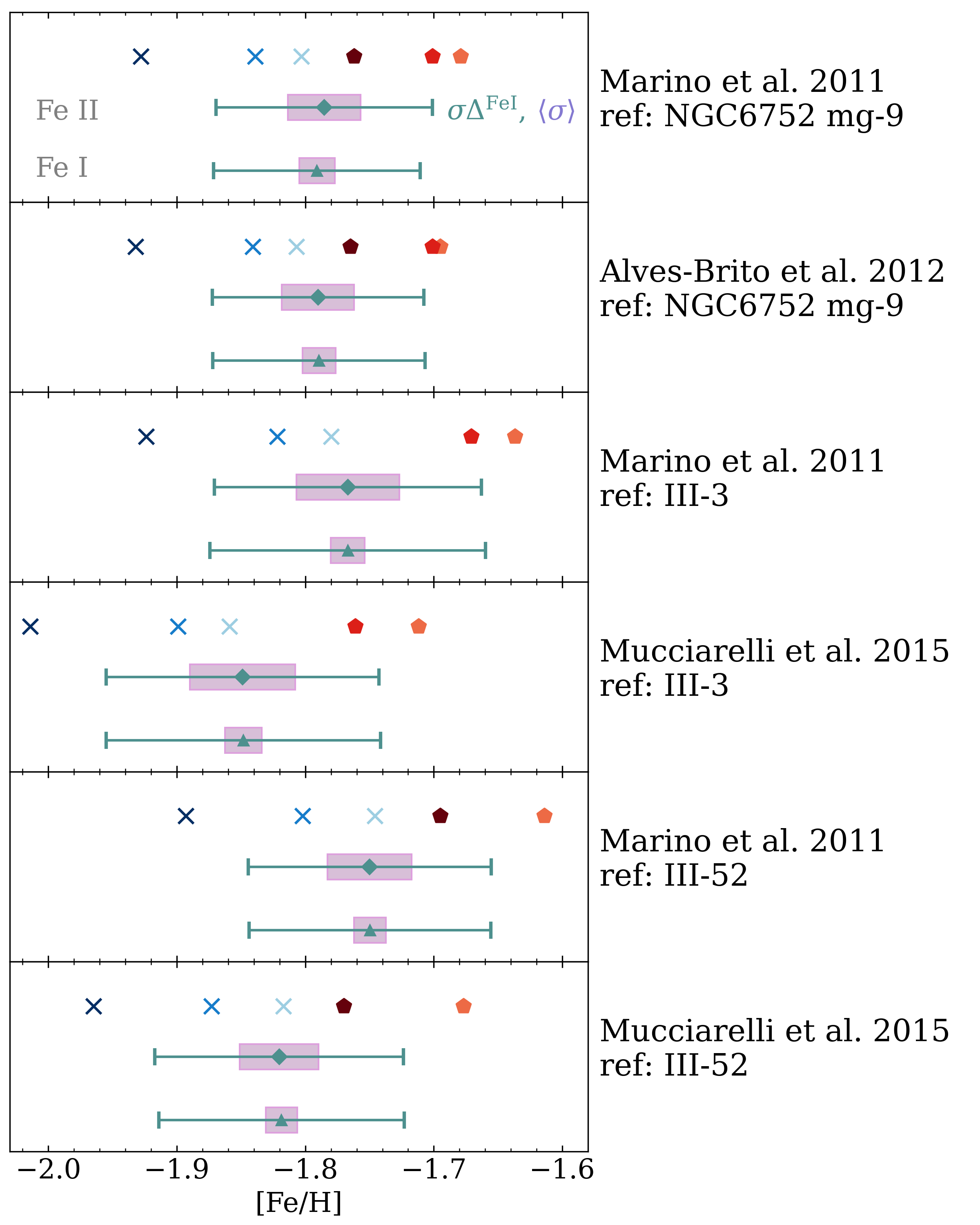}
    \caption{Each panel illustrates the spread in Fe given different initial values from the literature and different reference stars. The Fe II abundances are shown at the top of every panel, with the same colours and markers used in Figs \ref{fig:LaEu_Fe_relation} and \ref{fig:6_element_grid}. For both Fe I and Fe II abundances, the green error bar represents the 1$\sigma$ standard deviation of Fe for our program stars and the purple box on top of it is the average error for each of the stars. Fe I and Fe II abundances are almost identical (as $q^2$ has imposed ionisation balance), however Fe I has smaller errors for each star (i.e. smaller purple error bars) as more lines are measured. The results presented in this paper are from the top panel which uses stellar parameters from \protect\cite{Marino+2011} as starting values for the program stars with NGC6752 mg-9 as the reference star.}
    \label{fig:SP_spreads}
\end{figure}

We test the impact of different stellar parameters on iron abundances in Fig. \ref{fig:SP_spreads}. The text to the left of each panel describes the initial values and reference star used for abundance determinations. The top panel reflects abundances presented in this work, using the stellar parameters from \cite{Marino+2011} as starting values and the reference star, NGC6752 mg-9, from \cite{Yong+2013}. Additional models use \cite{Alves-Brito+2012} initial values for the program stars (except for the C star which uses \cite{Marino+2011} values) with NGC6752 mg-9 as the reference, \cite{Marino+2011} initial values with III-3 as the reference, \cite{Marino+2011} initial values for the program stars with \cite{Mucciarelli+2015} initial values for III-3 as the reference, all \cite{Marino+2011} initial values with III-52 as the reference, and \cite{Marino+2011} initial values for the program stars with \cite{Mucciarelli+2015} values for III-52 as the reference. In all panels, the abundances have been scaled by the metallicity of the reference star in order for each measurement to be in the same Fe range. However [Fe/H] abundances should not be compared between methods. If there was no abundance spread within M 22, we would expect that the green error bars of the Fe abundances would be the same length as the purple rectangles showing the size of the errors. However regardless of the model, the stellar parameters always converge on a sample of stars with a Fe spread (both Fe I and Fe II) greater than the errors expected for each star. Each iteration of determining stellar parameters arrives at slightly different stellar parameters depending on the reference star and the initial values provided. This is unsurprising as the documentation for $q^2$ explains that the final stellar parameters will be somewhat dependent on the initial parameters and parameter steps. However in each case, the scatter in Fe remains and the pattern between \textit{s}-process rich and poor stars is retained. The abundances derived using different initial guesses and reference stars are in agreement with those found from the fiducial model.

We perform additional tests by selecting a reference star from the Gaia benchmark sample \citep{Heiter+2015}. These stars have known radii, fluxes and distances which enable direct measurements of \teff\ and \logg\ from the Stefan-Boltzmann relation and Newton's law of gravity, respectively (\citealt{Heiter+2015}; \citealt{Jofre+2015}). We analyse MIKE spectra of the Gaia benchmark metal-poor red giant star HD 122563 in the same way as our program stars. NLTE and 3D-NLTE stellar parameters from \cite{Bergemann+2012} and \cite{Amarsi+2016}, respectively, were adopted and the M 22 program stars were analysed using HD 122563 as the reference. We find the relative abundances between \textit{s}-process poor and rich stars remain the same, but the errors are much larger due to the differences in parameters between HD 122563 and our reference and program stars (e.g. HD 122563 has [Fe/H] = -2.49 dex as compared to our most metal rich star, C, with [Fe/H] = -1.68 dex).

\subsubsection{Other non-detections}
\label{sec:other_non_detections}
Previous non-detections of iron abundance spreads have been due to limited sample size (e.g. \citealt{Cohen_1981}), abundance errors being larger than the intrinsic spread (e.g. \citealt{Anthony-Twarog+1995}) or techniques relating to the determination of stellar parameters (e.g. \citealt{Mucciarelli+2015}). A study that does not suffer from these issues is from \cite{Meszaros+2020}, which analysed 20 stars with a S/N>70 from the SDSS-IV APOGEE-2 survey. They estimate a higher [Fe/H] than previous studies for the cluster (-1.52 dex) with an average uncertainty of 0.09 dex. The Fe scatter is reported to be 0.112 dex, however, they say they cannot make strong statements about the intrinsic [Fe/H] scatter. Their results for the [Fe/H] scatter in $\omega$ Cen using 775 stars with a S/N > 70 are noticeably smaller than what is quoted in the literature; they detect a spread of 0.205 dex, whereas other studies report a value of $\sim$ 1.2 dex (e.g. \citealt{Norris_DaCosta1995}; \citealt{Johnson_Pilachowski2010}). Thus if the cluster with the largest heavy abundance variations known to date has a detected spread of 0.2 dex, then it is unsurprising that the "less extreme version" of this cluster would have an undetectable spread given their analysis techniques. \cite{Horta+2020} also used APOGEE abundances for M 22 in their work, but clipped around the [Fe/H] value from the 2010 edition of the Harris catalogue \cite{Harris_1996} to establish membership. \citep{Harris_1996} lists the metallicity of M 22 as [Fe/H] = -1.7, so the sample would have been skewed towards the \textit{s}-process rich population and it is understandable why no abundance spread was detected.

Photometric studies such as \cite{Anthony-Twarog+1995} or \cite{Monaco+2004} allow for a maximum range in Fe abundance variations of $\approx0.2$ dex, comparable to the minimum allowable spread presented in this work. We note that these studies are affected by difficulties in ascertaining the reddening of the cluster and do not account for C+N+O differences between the two populations which is estimated to be $\Delta$[C+N+O] $\approx$ 0.13 dex (\citealt{Marino+2011})\footnote{Note that this study measured only an upper limit to the N abundance for almost all the \textit{s}-poor stars, suggesting that the real difference could be larger.}. Theoretical works have shown that the T$_{\rm{eff}}$ of stars along the horizontal branch, as well as the mass loss along the RGB is affected by the CNO abundance (\citealt{Cassisi+2008}; \citealt{Ventura+2009}), which may therefore influence photometric results.

\subsection{The origin of M 22}
\label{sec:M 22_evolution}
Building an unambiguous evolutionary picture for a globular cluster is a difficult task, even more so for a cluster with a surplus of anomalous observations. This work, among many others, has demonstrated the existence of abundance spreads for almost all elements in M 22. Historically, this has led many to believe that, like $\omega$ Cen, it is a probable nucleus of a stripped dwarf galaxy (\citealt{Freeman1993}; \citealt{Bekki_Freeman2003}). However, there is a disparity between its chemical and dynamical histories. \cite{Massari+2019} used Gaia DR2 integrals of motion to assign GCs to accretion events to determine whether they formed within the Milky Way (MW). Their results suggest that M 22 is a member of the MW disc. It is difficult to envision a scenario where a merging event with the MW can deposit a cluster on such a disc-like orbit, thus in light of this and our current understanding of the chemical and dynamical state of M 22, we discuss three possible formation scenarios. We intend to investigate M 22's creation through the lens of Mg isotope ratios in future work.

\subsubsection{A nuclear star cluster?}
There is a historical precedent for M 22 to be labelled as a nuclear star cluster. As discussed in \cite{Hesser+1977}, \cite{LloydEvans_1978} and more recently in \cite{DaCosta+2009}, its similarities to $\omega$ Cen, especially in regards to its heavy element abundance spreads, indicate that these clusters share a similar origin. Furthering this claim, \cite{DaCosta_2016} discussed the theory that all clusters with substantial internal [Fe/H] spreads originated as nuclei of disrupted dwarf galaxies (for a recent review on nuclear star clusters (NSCs), see \citealt{Neumayer+2020}). Recently, \cite{Pfeffer+2021} claimed that metallicity variations, together with the extragalactic origin, are the requirements that a GC needs to be an NSC. Based on this hypothesis and the work from  \cite{Mucciarelli+2015} and \cite{Meszaros+2020}, they concluded that

"M 22 is not likely to host significant spreads in [Fe/H], hence it is not a candidate NSC". Our results, along with others listed in Table \ref{tab:publications}, provide strong evidence, from a variety of approaches, for the existence of an iron spread. However, the dynamical history of this cluster does pose some challenges for this formation scenario. Recent work identifying the origin of $\omega$ Cen through its dynamical history strongly suggests an extragalactic formation site (\citealt{Myeong+2019}), but M 22 does not share a similar history. In addition to \cite{Massari+2019}, \cite{Moreno+2021} also lists M 22 as a main disc cluster, with orbital parameters largely in agreement with previous estimations in the literature (\citealt{Gaia_Collaboration+2018}, \citealt{Bajkova_Bobylev2021} and the compilation by Holger Baumgardt\footnote{\url{https://people.smp.uq.edu.au/HolgerBaumgardt/globular/}} using the Galactic Model I of \citealt{Irrgang+2013}). Conversely, \cite{Callingham+2022} assigns an 85\% probability of M 22 belonging to the Gaia-Enceladus-Sausage, as found by their chemo-dynamical model.

\cite{Horta+2021} discussed the presence of a low metallicity accreted structure in the inner galaxy, however M 22's orbital energy and eccentricity (based on values from \citealt{Bajkova_Bobylev2021}) do not fit these criteria. Recently, \cite{Santistevan+2021} discussed the origin of metal-poor stars on prograde disc orbits 
using the FIRE-2 suite of cosmological zoom-in simulations. One of their findings was that a gas-rich merger could deposit a significant population of old metal-poor stars and gas into the host on the same orbital vector, which typically seeded/shaped the formation of a long-lived disc in the host. This resulted in metal-poor stars being preferentially on a prograde disc orbit. Future theoretical works should explore the possibility of a merging event at a low inclination to the disc depositing a GC with a disc-like orbit and with orbital energies and eccentricities similar to that of M 22. This would allow for the scenario where M 22 is a NSC. Furthermore, whether this accretion can occur before or after the MW disc has been established should also be investigated.

\subsubsection{The product of two merged clusters?}
$\omega$ Cen is not the only cluster that has frequently been compared to M 22. As discussed in \cite{Roederer+2011}, the heavy elements in NGC 1851 are reminiscent of the patterns observed in M 22. \cite{vandenBergh1996} and \cite{Campbell+2012} each discuss the possibility of merging clusters resulting in abundance spreads. \cite{Carretta+2010_1851} supported this idea for NGC 1851, measuring a [Fe/H] spread of 0.06-0.08 dex. This has been further investigated by \cite{Tautvaisiene+2021} who used the averaged A(C+N+O) values to make the case that NGC 1851 is composed of two clusters. Because of the chemical parallels between M 22 and NGC 1851 (see \citealt{Yong_Grundahl2008}, \citealt{Yong+2009} or \citealt{Carretta+2011} for NGC 1851), one could assume that M 22 is also the result of two merging clusters. Corroborating this idea, \cite{Lee_2016} decomposed M 22 into 5 different populations, thus satisfying the conventional first generation, second generation patterns observed in Galactic GCs.

Simulations of an anomalous GC being created through a merging event were discussed in \cite{Bekki_Yong2012} and \cite{Bekki_Tsujimoto2016}. They state that the merger must occur within a dwarf galaxy and the abundances of the resulting cluster can depend on the host galaxy's chemical evolution. In a galaxy with mass $\sim 10^{10}\rm{M}_{\odot}$, two massive GCs ($>3\times10^5\rm{M}_{\odot}$) can merge to form a single nucleus before its host is completely destroyed by a MW-like potential. A prediction of this scenario is that the resulting star cluster would be rotating and as such, M 22 has been confirmed to rotate by several studies (\citealt{Gaia_Collaboration+2018}; \citealt{Bianchini+2018}; \citealt{Vasiliev2019}; \citealt{Sollima+2019}; \citealt{Vasiliev_Baumgardt2021}). Additionally, structural differences between enriched populations can be achieved by varying the densities of the two clusters \citep{Gavagnin+2016}. \cite{Khoperskov+2018} and \cite{MastrobuonoBattisti+2019} conducted similar merging experiments but focused on clusters in the thick disc. They found that given a large, massive population of disc clusters, mergers, fly-bys and mass exchanges between GCs can occur over time. They calculate that a single merger event can occur each Gyr given a population of $\sim$100 GCs with initial masses of $10^7 \ \rm{M}_{\odot}$ in a galactic disc, with scale length and height similar to the current MW thick disc. In the context of M 22, this scenario negates the requirement for an accretion event to place M 22 on a disc-like orbit while also generating abundance variations. Cluster merging is not frequent enough to account for the growing number of Type II clusters with common chemical patterns (e.g. metal-rich stars are enhanced in \textit{s}-process elements and in their overall C+N+O abundances).

Despite both observational and theoretical support for this scenario, it is still unclear how the \textit{s}-process rich population (in this case, a fully-fledged GC prior to merging) acquired a surplus of these heavy elements in the first place. However, there are studies which have examined the role of AGBs in this scenario (\citealt{Shingles+2014}; \citealt{Straniero+2014}). Furthermore, as in Fig. \ref{fig:6_element_grid}, the differences in abundances between the two populations varies depending on nucleosynthetic site amongst other factors. Variations within each individual \textit{s}-process group (e.g. $\Delta^{\rm{Fe I}}$ variations of 0.06 and 0.04 dex for \textit{s}-process rich and poor populations respectively in Fig. \ref{fig:pop_diff}) also adds to the complexity of the scenario. How this chemical pattern can be established through a merging event is still an open question.

\subsubsection{A building block of the Milky Way?}

The stellar disc of the MW is a complex structure and its formation and evolution with regard to physical and dynamical processes continues to be debated (e.g. see \citealt{Katz+2021}). Large spectroscopic surveys have begun to dissect the metal poor tail of the disc (\citealt{Beers+2002}; \citealt{Kordopatis+2013}; \citealt{Li_Zhao2017}; \citealt{Sestito+2019}; \citealt{DiMatteo+2020}; \citealt{Fernandez-Alvar+2021}) and stars with kinematics akin to the thin and thick disc have been identified at metallicities ranging from [Fe/H] = -6 dex to -2 dex. Very recent work from \cite{Belokurov_Kravtsov2022} presents the discovery of the original in-situ component of the MW, dubbed \textit{Aurora}. These low metallicity ([Fe/H] $\lesssim$ -1.3 dex) stars were born before any coherent disc was established and reflect the chaotic pre-disc period of the MW's evolution. Within this component, they identified a small fraction of stars with chemical correlations reminiscent of typical GCs. Hence given M 22's metallicity range of $-2\lesssim$[Fe/H]$\lesssim-1.6$ and disc-like kinematics, we suggest that this cluster could have formed alongside this metal-poor component and be a remnant of the building blocks of the MW. Furthermore, \cite{Myeong+2022} determined that this component is enhanced in \textit{s}-process elements, indicating a common but not necessarily shared nucleosynthetic history between \textit{Aurora} and M 22. Thus, the assembly sequence of the MW could be encoded in the abundances of M 22 and be used to probe different nucleosynthesis processes during the earliest stage of our Galaxy's evolution.

The question of how M 22 survived the formation of the disc, and why it possesses such unusual chemical abundances remains unclear. The only other anomalous disc cluster as listed by \cite{Massari+2019} is NGC 7078 (M15; e.g. \citealt{Sneden+1997}; \citealt{Sneden+2000}; \citealt{Nardiello+2018}) and the rest of the GC population either has a typical Type I GC abundance pattern or is too poorly studied to comment on. If there once were more M 22-like clusters, they have since dissolved and have not been identified. N-rich stars, usually the markers of dissipated GCs, are most commonly discussed in the context of the halo (\citealt{Martell+2011}; \citealt{Martell+2016}; \citealt{Fernandez-Trincado+2017}; \citealt{Horta+2021_Nrich_stars}) or the bulge (\citealt{Schiavon+2017}; \citealt{Bekki2019}; \citealt{Fernandez-Trincado+2020}) but rarely in the disc. Furthermore, a bimodal \textit{s}-process element pattern is not apparent in the MW thick disc (e.g. \citealt{Tautvaisiene+2021_disc}).

A variation of the scenario commonly described for Type I GC formation could be adapted to the formation of M 22. 
In this scenario, GCs with two populations varying in light element abundances but not in heavy elements are formed through the ejecta of AGB stars, or a mixture of AGB ejecta and pristine gas with comparable abundances to the 1G (e.g. \citealt{Cottrell_DaCosta_1981}; \citealt{DErcole+2008}; \citealt{Renzini+2015}; \citealt{DAntona+2016}). Within this framework, two possible scenarios emerge that could lead to the abundances of M 22. One could assume that the \textit{s}-process poor population initially forms, then enriched AGB ejecta from these stars create the \textit{s}-process rich population. Two independent teams (\citealt{Straniero+2014}; \citealt{Shingles+2014}) examined this scenario and concluded that the \textit{s}-process abundance differences between the two stellar groups can be attributed to pollution from AGB stars with masses in the range 3-6 M$_{\odot}$. Alternatively, accreted gas enriched in \textit{s}-process elements could have also contributed to the \textit{s}-rich population. In both cases, a contribution from supernovae is also necessary to reproduce the Fe enhancement in the \textit{s}-process rich population of M 22. Further investigation into whether the light element variations can also be reproduced by this scenario is necessary. 
Three theoretical works which illustrate this process include \cite{Calura+2019}, \cite{McKenzie_Bekki2021b} and \cite{Lacchin+2021}. Each has had success in generating anomalous clusters which have a range in heavy element abundances.

Although this scenario can explain M 22's membership to the disc and has enough freedom in its formation scenario to explain the abundance patterns, the question of why this cluster is so unique when compared to the rest of the disc cluster population is something that requires further investigation.

\section{Conclusions}
\label{sec:conclusion}
The main result from our study is that high-precision differential abundance measurements have not only verified the metallicity spread in the cluster M 22, but also variations in each of its \textit{s}-process populations. The spectra analysed in this work are the highest quality ever obtained (in terms of spectral resolution and SNR), and the strictly differential nature of the analysis allows us to reach Fe uncertainties as low as $\sim$0.01 dex. Hence these are the most precise abundances ever obtained for this cluster. Our method addresses the concerns of NLTE calculations by using a well characterised reference star to compare our observations to. We take a boutique approach for determining equivalent widths by manually inspecting each line to honour and preserve the integrity of the data. The iron abundance spread persists even when using different reference stars and stellar parameters, determined both spectroscopically and photometrically. The iron spread (and \textit{s}-process element spread) reported here is in good agreement to that of \cite{Marino+2011}; the difference being that the errors have been reduced by a factor of $\sim$10. We find that across all measured $\alpha$ and iron peak elements, there is an average difference of $\approx$0.2 dex between the two populations, and a $\approx$0.5 dex difference for the \textit{s}-process elements. We find a positive correlation between all elements, with some combinations showing a bimodal separation whereas others have a gradual transition between the two \textit{s}-process groups. This dataset will be used to further examine the AGB scenario by measuring isotopic ratios of Mg which we hope will put further constraints on possible formation mechanisms. There is no clear evolutionary scenario that can account for both the chemical abundance patterns and the dynamical history discussed in other works. For now, M 22 remains yet another enigma of our Galaxy.

\section*{Acknowledgements}

The authors are grateful for the referee's thorough and fair assessment of this work.
We thank both Dr Kenji Bekki for the useful discussions of M 22 in the context of simulations and Dr Thomas Nordlander for his help in interpreting our differential stellar parameters and abundances in relation to NLTE effects. 

MM is supported by the Australian Government Research Training Program (RTP) Scholarship. 

This research was supported by the Australian Research Council Centre of Excellence for All Sky Astrophysics in 3 Dimensions (ASTRO 3D), through project number CE170100013.

I.U.R.\ acknowledges support from
the NASA Astrophysics Data Analysis Program (80NSSC21K0627), and 
the United States National Science Foundation
(PHY~14-30152: Physics Frontier Center/JINA-CEE, AST~1613536, and
AST~1815403/1815767).

This work has received funding from the European Research Council (ERC) under the European Union's Horizon 2020 research innovation programme (Grant Agreement ERC-StG 2016, No 716082 'GALFOR', PI: Milone, http://progetti.dfa.unipd.it/GALFOR), APM acknowledges support from MIUR through the FARE project R164RM93XW SEMPLICE (PI: Milone) and the PRIN program 2017Z2HSMF (PI: Bedin).

This research has made use of NASA's Astrophysics Data System and Bibliographic Services.
This research has made use of the SIMBAD database, operated at CDS, Strasbourg, France.
IRAF is distributed by the National Optical Astronomy Observatory, which is operated by the Association of Universities for Research in Astronomy (AURA) under a cooperative agreement with the National Science Foundation. This work has made use of data from the European Space Agency (ESA) mission
{\it Gaia} (\url{https://www.cosmos.esa.int/gaia}), processed by the {\it Gaia}
Data Processing and Analysis Consortium (DPAC,
\url{https://www.cosmos.esa.int/web/gaia/dpac/consortium}). Funding for the DPAC
has been provided by national institutions, in particular the institutions
participating in the {\it Gaia} Multilateral Agreement. This research has made use of the AstroBetter blog and wiki.

\section*{Data Availability}

Data are available from ESO website with program ID "095.D-0027(A).



\bibliographystyle{mnras}
\bibliography{refs}



\appendix

\section{Radial Velocities}
We provide the heliocentric radial velocities for our stars along with a comparison to the literature in Table \ref{tab:rvs}. This is only a subset of published measurements and not an exhaustive list of values in the literature. Our values are in good agreement to those from Gaia DR3 \citep{Gaia_DR3_2022}.

\begin{table*}
\centering
    \caption{Heliocentric radial velocities as measured by our study along with a sample of literature values.}
    \label{tab:rvs}
\begin{tabular}{lcccccc}
\hline
 & This work & \cite{Peterson_Cudworth1994} & \cite{Cote+1996} & \cite{Marino+2011} & \cite{Alves-Brito+2012} & \cite{Gaia_DR3_2022} \\ \hline
C      & -149.05 &  -      & -154.32 &    -    &    -   & -151.40 \\
III-3  & -146.48 & -147.46 & -148.26 & -148.16 & -148.6 & -146.94 \\
III-14 & -150.80 & -152.72 & -150.98 &  -   & -150.2 & -149.78 \\
III-15 & -146.46 & -144.45  & -147.47 &   -     & -148.3 & -145.58 \\
III-52 & -150.66 & -150.40 & -151.95 & -153.22 & -148.8 & -149.95 \\
IV-102 & -137.87 & -142.29 & -140.71 &    -    & -149.1 & -140.14 \\ \hline
\end{tabular}
\end{table*}

\section{Details on Routine for EValuating and Inspecting Equivalent Widths (\textsc{review})}
\label{appendix:REVIEW}
We developed our own code to determine the equivalent widths (EWs) of lines in our program stars (available from \url{https://github.com/madeleine-mckenzie/REvIEW}). One advantage of this code over others currently in use is that it does not require additional package installs that are not regularly used in the python language.
Synthetic spectra were generated to span a range of EWs, possible locations of a line and deviations from a normalised continuum.

Normalised line depths spanned from 0.06 to 0.7, line widths from 0.03 to 0.15 \AA{}, line positions between $\pm$ 0.2 \AA{} from the predicted location in the line list, and continuum values $\pm$ 0.02 from the normalised continuum. These ranges were chosen to be consistent with the variation found in our target stars. Up to 4 Gaussian lines are placed within the 1.2 \AA{} window in which we fit our lines to account for blended lines and various degrees of random noise are added to the spectra to allow for EW measurements for spectra at lower S/N regions.

We train a feed forward neural network (FFNN) on the synthetic spectra to validate the EWs recovered through curve\_fit. The FFNN is trained with the ReLU activation function, a maximum of 500 iterations, $10^{-4}$ tolerance, and 10\% holdout. We do a hyperparameter search over layers, neurons, and the L2 penalty and find the best hyperparameters to be 4 layers, 300 neurons, and L2 penalty of $10^{-5}$. Our best hyperparameters are picked based upon a labelled real data set as opposed to the synthetic data set. We note that this does cause data leakage but as we are only testing the validity of our results with the FFNN and not using the actual values, this data leakage should not affect our final results.

Our real data are not always sampled in the same way as our synthetic spectra, and thus we interpolate over each line using a cubic spline. The choice of interpolation function has no statistical impact on the determination of the EW. Although interpolation of the data is necessary to account for different
pixel sizes and FWHM.

\begin{figure}
    \includegraphics[width=\columnwidth]{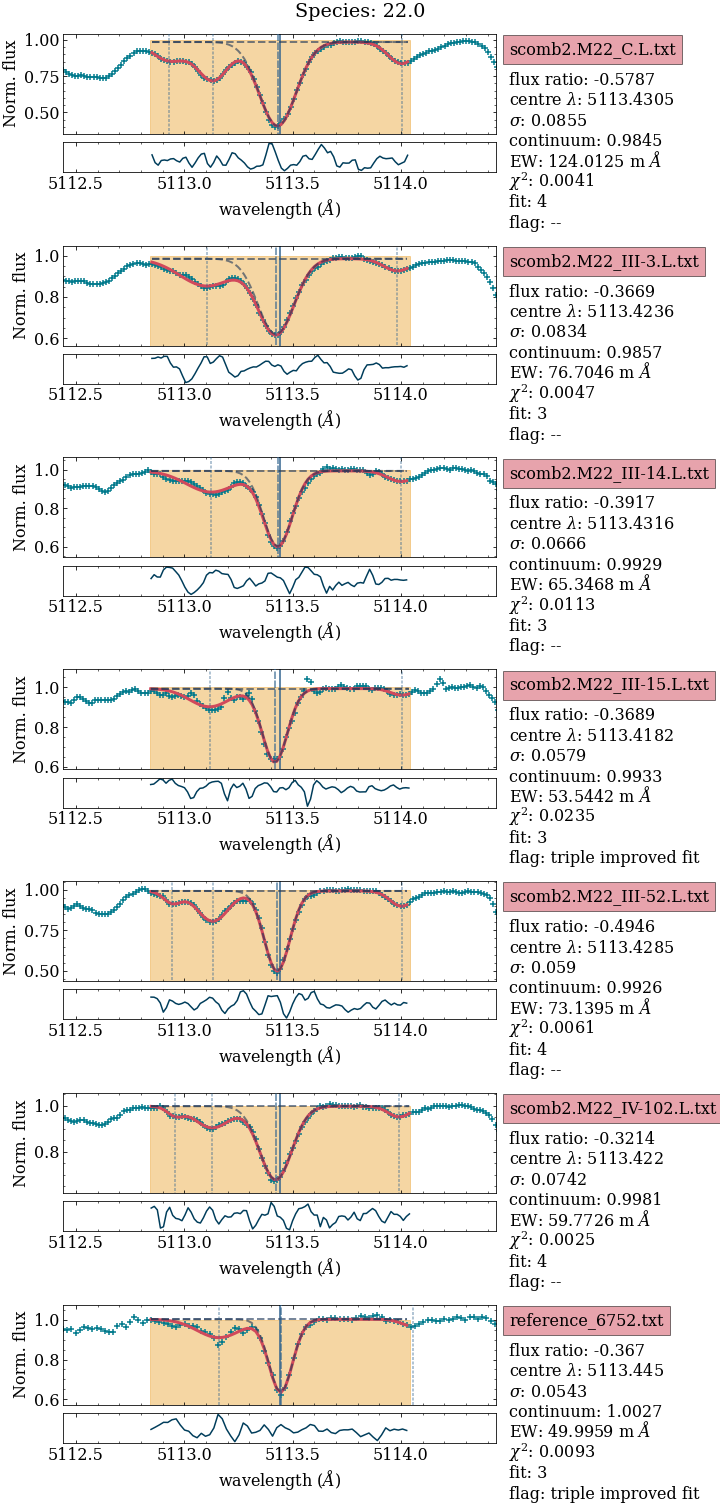}
    \caption{An example output from our equivalent width fitting code \textsc{review}. An image is generated for each line in the line list so that the quality of fitting can be determined by eye. We plot a titanium line with additional absorption features to the blue and red. For each panel, the spectra (green crosses) has been fit by a variable number of Gaussians within the yellow region, and we can confirm that both the line and the continuum has been fitted correctly. In higher S/N spectra, the algorithm identifies four lines in this region without the need for a detailed line list. Where there is more noise, it reverts to three Gaussians to model the line.
    }
    \label{fig:REvIEW_output}
\end{figure}

It does slightly decrease the accuracy of the neural network ($\pm$2 m\AA{} for our validation data). 
We provide an example of the output of one star in Fig. \ref{fig:REvIEW_output}. More detailed explanations of each of the returned quantities is provided on the github page.

\begin{figure}
    \includegraphics[width=\columnwidth]{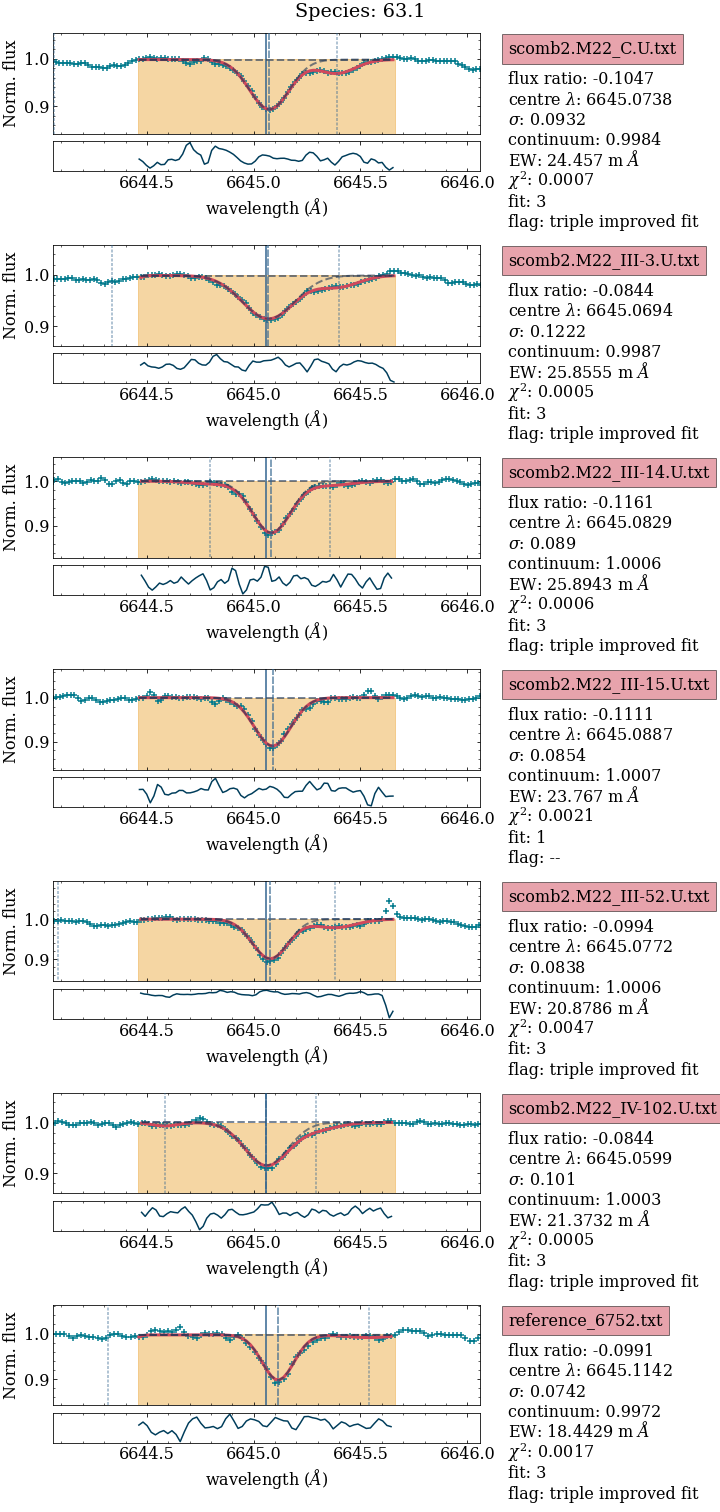}
    \caption{The same as Fig. \ref{fig:REvIEW_output} but for the \textit{r}-process element Eu. The code can still correctly measure the line, despite the blended iron line to the right.
    }
    \label{fig:REvIEW_output2}
\end{figure}

\section{Stellar parameters from the literature}
\label{sec:lit_stelar_params}
See Table \ref{tab:lit_stelar_params} for a compilation of literature values for the stellar parameters of our sample stars. We list the Fe I values for \cite{Mucciarelli+2015}.

\begin{table*}
\centering
    \caption{Literature values of stellar parameters for our program stars.}
    \label{tab:lit_stelar_params}
\begin{tabular}{lcccccc}
\hline
\textbf{Publication}                                 & \textbf{C}      & \textbf{III-3  }           & \textbf{III-14} & \textbf{III-15 }           & \textbf{III-52}            & \textbf{IV-102 }  \\ \hline \hline
\multicolumn{7}{c}{T$_{\rm{eff}}$ (K)}    \\ \hline 
\cite{Peterson_1980}& -      & 4000              & -      & -                 & -                 & -        \\
\cite{Gratton_1982}& -      & -                 & -      & -                 & 4150              & -        \\
\cite{Pilachowski+1982}& -      & 4100              & -      & -                 & -                 & 3900     \\
\cite{Frogel+1983}& -      & 4165              & 4062   & -                 & 4192              & 4086     \\
\cite{Wallerstein+1987}& -      & 4150              & -      & -                 & - & 4050     \\
\cite{Gratton_Ortolani1989}& -      &                   & -      & -                 & 4192              & -        \\
\cite{Brown+1990}& -      & 4500              & -      & -                 & 4200              & 4400     \\
\cite{Brown_Wallerstein1992}& -      & 4165              & -      & -                 & 4200              & 4090     \\
\cite{Marino+2009}& -      & 3990              & -      & -                 & 4100              & -        \\
\cite{Marino+2011}& 3960   & 4000              & 4030   & 4070              & 4075              & 4020     \\
\cite{Alves-Brito+2012}& -      & 3952              & 3919   & 4055              & 4124              & 3974     \\
\cite{Mucciarelli+2015} (method 1) & -      & 3910              & -    & -   & 4060                             & -        \\
\cite{Mucciarelli+2015} (method 2)                 & -      & 3960              & -   & -   & 4070                 & -        \\
\cite{Mucciarelli+2015} (method 3) & -      & 3992              & -      & - & 3986                  & -        \\
This work                                   & 3912 & 4041 & 4038 & 4136	& 4100 & 4043 \\
Average                                     & 3936 & 4071 & 4012 & 4087 & 4121 & 4070 \\\hline
                                            \multicolumn{7}{c}{log \textit{g} (cm s$^{-2}$)}                                                              \\\hline
\cite{Gratton_1982}&-        &-                   &-        &-                   & 0.3               &-          \\
\cite{Pilachowski+1982}&-        & 0.7               &-        &-                   &-                   & 0.9      \\
\cite{Frogel+1983}&-        & 0.5               & 0.5    &-                   & 0.8               & 0.5      \\
\cite{Wallerstein+1987}&-        & 0.5               &-        &-                   &-                   & 0.5      \\
\cite{Gratton_Ortolani1989}&-        &-                   &-        &-                   & 0.8               &-          \\
\cite{Brown+1990}&-        & 0.7               &-        &-                   & 0.8               & 0.5      \\
\cite{Brown_Wallerstein1992}&-       & 0.5               &-        &-                   & 0.8               & 0.5      \\
\cite{Marino+2009}&-        & 0.2               &-        &-                   & 0.67              &-          \\
\cite{Marino+2011}& 0.3    & 0.3               & 0.35   & 0.4               & 0.6               & 0.2      \\
\cite{Alves-Brito+2012}&-        & 0.43              & 0.41   & 0.59              & 0.72              & 0.43     \\
\cite{Mucciarelli+2015} (method 1)                 &-        & 0.4               &-        &-                   & 0.57              &-          \\
\cite{Mucciarelli+2015} (method 2)                 &-        & 0.34              &-        &-                   & 0.63              &-          \\
\cite{Mucciarelli+2015} (method 3) &-        & 0.36              &-        &-                   & 0.65              &-          \\
This work                                   & 0.105	& 0.25 & 0.12 & 0.45 & 0.51 & 0.1 \\
Average                                     & 0.20 & 0.43 & 0.35 & 0.48 & 0.65 & 0.45 \\\hline
                                            \multicolumn{7}{c}{[Fe/H]}                                                        \\\hline
\cite{Peterson_1980}&-        & -1.62             &-        &-                   &-                   &-          \\
\cite{Gratton_1982}&-        &-                   &-        &-                   & -1.89             &-          \\
\cite{Pilachowski+1982}&-        & -1.35             &-        &-                   &-                   & -1.7     \\
\cite{Wallerstein+1987}&-        & -1.5              &-        &-                   &-                   & -1.7     \\
\cite{Gratton_Ortolani1989}&-        & -                  &-        &-                   & -1.7              &-          \\
\cite{Brown+1990}&-        & -1.7              &-        &-                   & -1.6              & -1.9     \\
\cite{Brown_Wallerstein1992}&-        & -1.55             &-        &-                   & -1.56              & -1.78    \\
\cite{Marino+2009}&-        & -1.66             &-        &-                   & -1.62             &-          \\
\cite{Marino+2011}& -1.60  & -1.72             & -1.82  & -1.82             & -1.63             & -1.97    \\
\cite{Alves-Brito+2012}&-        & -1.62             & -1.64  & -1.72             & -1.54             & -1.87    \\
\cite{Mucciarelli+2015} (method 1) &-        & -1.84             &-        &-                   & -1.72             &-          \\
\cite{Mucciarelli+2015} (method 2)&-        & -1.8              &-        &-                   & -1.7              &-          \\
\cite{Mucciarelli+2015} (method 3) &-        & -1.8              &-        &-                   & -1.68             &-          \\
This work & -1.70 & -1.78 & -1.87 & -1.83 & -1.71 & -1.97   \\
Average & -1.65	& -1.66 & -1.78 & -1.79 & -1.67 & -1.84 \\ \hline \hline
\end{tabular}
\end{table*}


\bsp	
\label{lastpage}
\end{document}